\documentclass[12pt]{article}
\usepackage{graphicx}
\usepackage{epsf}
\newcommand{\be}[1]{\begin{equation}\label{#1}}
\newcommand{\ee}{\end{equation}}
\newcommand{\bea}[1]{\begin{eqnarray}\label{#1}}
\newcommand{\eea}{\end{eqnarray}}
\newcommand{\gl}[1]{Eq.\,(\ref{#1})}

\newcommand{\micron}{\,\mu m}

\newcommand{\CC}{{\rm CC}}
\newcommand{\CB}{{\rm CB}}
\newcommand{\Sig}{{\rm Sig}}
\newcommand{\et}{{\it et al.}, }
\begin{document}
\vspace*{2cm}
\begin{center}
{\LARGE \bf  A mathematical model for the germinal
center morphology and affinity maturation}\\
\vspace{1cm}
Michael Meyer-Hermann\\
\vspace{1cm}
%
Institut f\"ur Theoretische Physik, TU Dresden,
D-01062 Dresden, Germany\\
E-Mail: meyer-hermann@physik.tu-dresden.de\\
\end{center}

\vspace*{2cm}

\noindent{\bf Abstract:}
During germinal center reactions the appearance of two specific zones 
is observed: the dark and the light zone. 
Up to now, the origin and function of these zones are
poorly understood. 
In the framework of a stochastic and discrete model
several possible pathways of zone development
during germinal center reactions are investigated. 
The importance of the zones in the germinal center 
for affinity maturation, i.e.~the process
of antibody optimization is discussed.
\vspace*{\fill}
\eject
\newpage

\section{Introduction}

Germinal centers (GC) are an important part of the humoral
immune response. They develop after the activation of
B-cells by antigens. Such activated B-cells migrate into the follicular
system where they begin a monoclonal expansion in the environment
of follicular dendritic cells (FDC). 
This phase of pure centroblast multiplication is followed by a phase of
intense hypermutation and proliferation which leads to a larger repertoire
of different centroblast types. This seems to be the
basis of the affinity maturation process, i.e.~the optimization
process of the antibodies with respect to a given antigen
during germinal center reactions.
Antigen fragments are presented on the surface of the FDCs which
enables antibody presenting cells to interact with them. It is widely
believed that centroblasts do not present antibodies
(Han \et 1997).
Accordingly, the selection process must begin 
after the differentiation of centroblasts to centrocytes
(the initiation of this differentiation process is still unclear)
which do neither proliferate nor mutate but which present
antibodies. The centrocytes have initiated a process of apoptosis,
so that they have a finite life time 
(Liu \et 1989; Liu \et 1994).
They have to be rescued from apoptosis through
an interaction with the antigen fragments on the FDCs and
with T-helper cells 
(Brandtzaeg, 1996; Tew \et 1997; Hollmann \& Gerdes, 1999; Hur \et 2000; Eijk \et 2001).
This at least two step selection process 
(Lindhout \et 1997)
primarily depends on the affinity of the tested antibody and
the antigen. Therefore, the affinity maturation during GC
reactions rests on two major pillars:
the hypermutation of centroblasts and the affinity dependent
antigen-specific selection of centrocytes.

The morphology of the GCs is very specific and shows properties
which are characteristic for different stages of the reaction
(Liu \et 1991; Camacho \et 1998).
In a first stage -- the phase of monoclonal expansion --
a system of FDCs is
continuously filled with proliferating centroblasts. 
Then the centroblasts begin
to differentiate into centrocytes and the selection process begins.
In this second phase the two types of B-cells, centroblasts
and centrocytes, form two zones 
(Nossal, 1991):
the dark zone which is dominated by proliferating and hypermutating
centroblasts, and the light zone containing centrocytes and FDCs.
These two zones are stable for at least some days 
and may then disappear.
In the third morphological phase the centroblasts
and centrocytes are homogeneously distributed over the GC
(Camacho \et 1998).
The number of cells is now continuously declining 
until the end of the GC reaction.
The total GC life time is about $21$ days 
(Jacob \et 1991; Liu \et 1991; Kelsoe, 1996).

Up to now, the role and the reason for the appearance 
of the dark and the light zone have not been resolved. These problems
are here investigated in the framework of a stochastic 2+1-dimensional
space-time model. Until now, only dynamical models for the GC reaction
without resolution of spatial aspects have been developed 
(Oprea \& Perelson, 1996, 1997; Rundell \et 1998; Kesmir \& de Boer, 1999;
Oprea \et 2000; Meyer-Hermann \et 2001).
However, for the analysis of the GC zones this is essential.
In the present article several possible dynamic pathways and requirements for
the development of dark and light zones are discussed.
The stability of the zones, i.e.~the duration of centroblast-centrocyte separation,
and their role within the affinity maturation process are analysed.

It is very important
to carefully analyse the existing experimental
knowledge. Therefore, the discussion of the model assumptions
(see Sec.~\ref{modell})
will frequently refer to the experimental situation, and the model
parameters will be determined directly or indirectly using corresponding data.
The section on the model development is divided into three parts:
A short review of the shape space concept for the representation
of antibody types (Sec.~\ref{shapespace}), 
a discussion of the dynamical properties of the GC during different
stages of the reaction (Sec.~\ref{phases}), and the introduction of the
spatial aspects, i.e.~cell movements and cell interactions
(Sec.~\ref{lattice}). Naturally, these subsections are interlinked,
so that some repetition is unavoidable.
The dynamical properties and the local cell interactions
lean upon a deterministic model developed before 
(Meyer-Hermann \et 2001).
A presentation of the results follows in Sec.~\ref{results}
including a discussion of necessary and sufficient requirements for the
development of dark zones (Sec.~\ref{origin}), of the implications
of dark zones for affinity maturation (Sec.~\ref{duration} and \ref{quality}), 
and of the robustness of the results with respect to
different physiological quantities entering the model (Sec.~\ref{robust}). 
Finally, the results are summarized and evaluated (Sec.~\ref{discuss}).


\section{Model development}
\label{modell}

A model describing morphological properties of the GC reaction
needs to have three basic aspects: the representation of antibodies,
the dynamics and the interaction between the cells,
and their spatial distribution. 
These aspects already define the structure of this section. 
It is worth emphasizing that this section is not only devoted
to the formulation of postulates, but also to give an experimental 
motivation for the
assumptions and to determine the physiological quantities which enter
the model. 
A careful discussion of the physiological parameters builds
the fundament of the model stability and of its relation to real GCs.
The resulting parameter values are summarized in Table \ref{parameter}.

\subsection{Representation of antibody and antigen phenotype}
\label{shapespace}

In order to describe the affinity maturation process during GC
reactions, the antibodies encoded by B-cells as well as
antigens are represented with the help of the shape space concept
(Perelson \& Oster, 1979).
A $D$-dimensional finite size lattice is defined
in which each point corresponds to one antibody phenotype (accordingly,
the shape space does not provide a representation of the genetic
variability of antibodies). A phenotypically relevant mutation is
represented by a jump to a nearest-neighbor point in the shape space,
which means that a single mutation of the B-cell does not lead to
a random change of the antibody phenotype.

The antigen is represented in the same shape space using the
assumption of optimal complementarity, i.e.~it exists a corresponding
type of antibody having a maximal affinity. The antigen is represented
by the point in the shape space, which corresponds to this
optimal antibody.

The representation of mutations
by jumps to neighbor points in the shape space allows
the affinity of antibodies of mutating B-cells to be enhanced successively,
which is believed to happen during GC reactions. 
This property may be called {\it affinity neighborhood} and 
means that antibodies on neighboring points in the shape space do not have
drastically different affinities with respect to a given antigen.
The relevance of possibly existing key mutations that
imply considerable changes of the affinity was discussed
before (see Meyer-Hermann \et 2001).

This concept allows a quantitative formulation of a smooth affinity function
on the shape space, describing the affinity of a given antibody to
a given antigen. Denoting the
antibody type and the antigen by the shape space points
$\phi$ and $\phi^*$, respectively, one can define the affinity
weight function
\be{affinity}
a(\phi,\phi^*) \;=\; \exp\left(-\frac{||\phi-\phi^*||^\eta}{\Gamma^\eta}\right)
\quad,
\ee
where $\Gamma\approx 2.8$ is the width of the affinity weight function,
$||\cdot ||$ denotes the Euclidean metric, and $\eta=2$ is the exponent
of the distance in the shape space.
These values were determined from existing data of affinity
enhancement in dependence on the number of observed mutations
(for details see Meyer-Hermann \et 2001,
Fig.~2).

In a physiological GC reaction we do not only expect to find different
antigen fragments but also different antigens. Nevertheless, in
experiments the immunization is often performed by the injection of
one specific antigen. Therefore, in the present model 
we assume only one antigen to be present during the GC reaction. 
However, it may be interesting
to investigate the dynamical changes of GC reaction with more
than one antigen.

\subsection{Germinal center phases}
\label{phases}

There exist many different pictures describing the stages of a 
GC reaction, both from the perspective of dynamical interactions and of
morphological properties. Here, a rather general point of view is adopted 
which is based on widely accepted observations and 
interpretations of the GC. We hope that this enhances the validity of
the model implications.

The whole GC reaction is initiated by the migration
of antigen-activated B-cells into the follicle system. The model
description starts when the activated B-cells are present
in the environment of antigen presenting FDCs in the follicle system.
The affinity of these cells to the antigen is assumed to be low
but non-vanishing (as they were activated). 
In the model this is represented
by a typical distance of the B-cells from the antigen in the shape
space between $5$ and $10$, i.e.~the B-cells have to hypermutate
at least $5$ times in order to find the optimal antibody type
with respect to the antigen. This number is in accordance
with the one observed in experiment 
(K\"uppers \et 1993; Wedemayer \et 1997).
Note that here only the hypermutations
of phenotypical relevance are counted. 

\subsubsection{The phase of monoclonal expansion}

The activated B-cells (centroblasts) start to multiply in the 
presence of the FDCs. 
This process of intense expansion -- 
the proliferation rate is about $1/(6hr)$ 
(Hanna, 1964; Zhang \et 1988; Liu \et 1991)
-- is believed to be monoclonal 
(Jacob \et 1993; McHeyzer-Williams \et 1993; Pascual \et 1994a; Han \et 1995a),
so that no somatic hypermutations occur in this stage of the GC
reaction, and the antibody type encoded by the centroblasts remains 
unchanged. 
Furthermore, the expanding centroblasts do not seem to present
antibodies on their surface 
(Han \et 1997)
-- at least in this first phase
(this has recently been questioned 
(Berek, 2001)).
As a consequence, in the model 
no interaction with antigen fragments or
T-helper cells is provided during this phase. 

GCs develop oligoclonally 
(Kroese \et 1987; Liu \et 1991; Jacob \et 1991; K\"uppers \et 1993),
i.e.~the number of initial centroblasts is small. 
It has been shown that after three days of centroblast expansion
all B-cells in average stem from about three to six seeder cells.
Within three days (this is the duration of this first phase 
(Liu \et 1991))
the FDC network is completely filled with centroblasts
(Camacho \et 1998).
The volume occupied by centroblasts even exceeds
the FDC network. The total number of centroblasts reaches about $12000$
cells in this stage of the reaction. 

%

\subsubsection{Early optimization phase}

The phase of B-cell selection by interaction of antibody presenting
centrocytes with antigen fragments on the FDCs 
is characterized by the following additional dynamical properties:

\paragraph{Somatic hypermutation}
The centroblasts continue to proliferate at the same rate. 
Additionally, they have a high probability of somatic hypermutations
of $m=0.5$ 
(Berek \& Milstein, 1987; Nossal, 1991)
leading to a higher variability of antibody types
in the GC. This property provides the basis of the affinity maturation
process and is incorporated into the model by a random probability
of $m=0.5$ for each dividing centroblast to switch to a randomly chosen
neighbor point in the shape space. Consequently, in average one of the two
new centroblasts emerging from a cell division will correspond to 
a changed antibody.

\paragraph{Differentiation to centrocytes}
The centroblasts differentiate into centrocytes 
(Liu \et 1991; MacLennan, 1994; Choe \et 2000).
The initiation
of this differentiation process is still unclear. One may suspect a
centroblast differentiation after a fixed lifetime or an unknown
signal of unknown provenience. Several options for the initiation
of the differentiation process will be discussed in the 
framework of the model.
The process itself is assumed to take some time, i.e.~a centroblast
in the state of differentiation will become a centrocyte with a
rate $g_1$.
This rate determines the speed of the whole GC reaction 
(Meyer-Hermann, 2001)
and has been observed in experiment. At a certain stage of the
GC reaction the proliferating cells were labeled and it was
found that after $7$ hours these cells were present in the light
zone in a non-proliferating stage 
(Liu \et 1991; Berek, 2001),
i.e.~they differentiated to centrocytes 
and moved into the light zone within $7$ hours. Therefore, the
differentiation rate has to be slightly larger than $1/(7hr)$. 
On the other hand, it should not differ drastically from the proliferation
rate as otherwise the total cell population would either die out
or explode on a small time scale 
(Meyer-Hermann, 2001).
Taking these arguments together one is led to a centroblast to centrocyte
differentiation rate of approximately $g_1=1/(6hr)$.

\paragraph{Apoptosis of centrocytes}
Unbound centrocytes have initiated an apoptotic process.
If apoptosis is not
suppressed in time the centrocytes will die. The typical life
time of centrocytes has been estimated to be $6$ to $16$ hours
(Liu \et 1994).
Correspondingly, the rate of centrocyte death $z$ is 
chosen in this range. A value of $z=1/(7hr)$ turns out to be
most advantageous for the affinity maturation.
Dead cells are rapidly engulfed by macrophages 
so that in the model they are removed from the lattice rapidly.
The speed of this process has a marginal impact 
on the space available for the movement of the 
remaining cells.

\paragraph{Centrocyte-FDC-interaction}
As the GC contains more and more centrocytes due to the
centroblast differentiation process during this phase, 
the interaction of the presented
antibodies with the antigen fragments on the FDCs has to be taken into
consideration. The inhibition of centrocyte apoptosis is believed
to depend on this interaction process as well as on the interaction 
with T-helper cells 
(Lindhout \et 1995; Brandtzaeg, 1996; Tew \et 1997; Hollmann \& Gerdes, 1999;
Hur \et 2000; van Eijk \et 2001).
Conversely, also the
acceleration of apoptosis of non-binding centrocytes has
been discussed 
(Hollmann \& Gerdes, 1999).
In the framework of such regulation processes, 
the role of CD40-CD40L-interaction, Fas-FasL-interaction, 
and bcl-2-expression are currently under investigation 
(Han \et 1997; Choe \& Choi, 1998; Choe \et 2000; van Eijk \et 2001; Siepmann \et 2001).
We do not propose a solution to this puzzle in the model,
but we effectively incorporate the inhibition of apoptosis.

In order to allow affinity maturation
the interaction of centrocytes with FDCs 
is assumed to depend on the affinity
of antibody and antigen 
(Liu \et 1989; Koopman \et 1997; Radmacher \et 1998).
In the model this is represented
by the affinity weight function \gl{affinity}: Each time 
a centrocyte is in local contact with an FDC, i.e.~they 
are direct neighbors on the spatial lattice,
it will bind to the antigen with a probability $a(\phi,\phi^*)$ that
is given by the affinity of the antibody presented by the centrocyte
to the antigen fragment presented by the FDC.
An effective binding rate for centrocytes and FDCs is 
implicitly introduced
using corresponding rules for the interaction
(see Sec.~\ref{represent} {\it Centrocytes}). This binding rate 
basically depends 
on the centrocyte diffusion constant $D_{\rm CC}$ and
on the cell distribution around the centrocyte.

Once a centrocyte is bound, it remains bound to the FDC for a certain time. 
The above mentioned rescue processes are thought to take
place during this time.
Unfortunatly, the duration of binding is not known
experimentally. However, the observation that
the process of apoptosis of a centrocyte is stopped within $2$ hours
(Lindhout \et 1995; van Eijk \& de Groot, 1999)
may provide a valuable hint:
Assuming that the inhibition of apoptosis is done
during the binding of centrocyte and FDC and that the centrocytes
dissociate from the FDC afterwards, the rate of centrocyte-FDC
dissociation becomes $g_2=1/(2hr)$.

\paragraph{Recycling of centrocytes}
Positively selected centrocytes, i.e.~B-cells encoding antibodies of
high affinity to the antigen, may reenter the state of proliferation
and mutation. This possibility is the frequently
discussed recycling hypothesis 
(Kepler \& Perelson, 1993; Han \et 1995a; Oprea \et 2000; Meyer-Hermann \et 2001),
which has not been proven in experiment until today.
It has to be clarified what is explicitly meant by recycling. 
In some definitions of the recycling process the recycled cells
do not only restart to proliferate and to mutate but also
return to the dark zone 
(Kelsoe, 1996; Han \et 1997).
In the present model a more modest variant is adopted, 
i.e.~recycling solely denotes the
possibility of reentering the state of proliferation and mutation --
independently of where this happens and where these cells tend to
go. Therefore, our model does not exclude proliferating cells in the 
light zone, and it will be discussed if a reentry into the dark zone is 
possible or even necessary.

\paragraph{Differentiation into plasma- and memory-cells}
During the phase of early optimization no production of plasma- 
and memory-cells is provided by the model (see Sec.~\ref{outputphase}),
i.e.~all positively selected centrocytes are recycled.

\paragraph{Signal molecules}
All processes described so far are based on
the production of signal molecules which lead to a {\it local} interaction
between the cells, i.e.~each signal is related to an interaction with a very
specifc B-cell. This especially is the case for the inhibition
of apoptosis in positively selected centrocytes as well as for the
differentiation path of rescued B-cells 
(into plasma- or memory-cells or back into a stage of proliferation).
Such local interaction processes are assumed to be necessary for affinity maturation, as
it is difficult to imagine a non-local interaction rescuing
specific B-cells of high affinity to the antigen. Therefore, we assume
in our model that these interactions take place during the
binding of centrocytes to FDCs. For this reason no
signal molecule has to be specified explicitly, since it is completely
sufficient to incorporate the corresponding functional
answer of the cells into the model.

Nevertheless, one should be aware that the centrocytes take the
antigen fragments with them, i.e.~B-cells that have successfully
interacted with the FDCs are recognizable in this way. Therefore,
a non-local interaction with a diffusing signal molecule (which
may be secreted for example by the FDCs) which
targets specific B-cells with bound antigen fragments may exist. 
Such a specific non-local interaction may be in accordance with
affinity maturation if a mechanism is utilized which recognizes
B-cells transporting antigen fragments. This case is not
covered by the model assumption of local interactions.
However, the efficiency of such a non-local interaction
process would be smaller compared to a local interaction:
If a high affinity B-cell is detected at the FDCs, this information
is lost through its dissociation from the FDC, and the same cell
must be detected for a second time by signal molecules.
This is an unreasonable procedure, especially
for the inhibition of apoptosis which is a time critical process. 

The situation is different for the differentiation of
centroblasts to centrocytes. 
Adopting the reasonable assumption that all
centroblasts will differentiate into centrocytes independently
of their affinity to the antigen
there exists no obvious reason for a local interaction. 
On the other hand, the presence
of FDCs and T-cells is believed to be important for the 
differentiation process of centroblasts 
(Dubois \et 1999a, 1999b; Choe \et 2000).
So if FDCs or T-cells trigger the differentiation of centroblasts
but not by a local interaction with single B-cells, 
a non-local interaction becomes obligatory.
In order to allow a non-local interaction in the model
we incorporate a diffusing signal molecule 
(see Sec.~\ref{represent}) which triggers the centroblast differentiation,
is produced by the FDCs, and is bound by the centroblasts.
This feature should not be considered an assumption but an option of the model. 
Alternative pathways of differentiation will be discussed.

\subsubsection{The phase of late optimization and output production}
\label{outputphase}

The duration of the early optimization phase $\Delta t_2$
determines the starting
point of the phase of late optimization and output production,
which differs from the early optimization
phase solely by an additional pathway of differentiation for
positively selected centrocytes.
It is widely believed that the interaction between centrocytes and FDCs
may involve a local differentiation signal which enables the centrocytes
to further differentiate into plasma or memory cells (shortly denoted
as {\it output cells} in the following) or -- as discussed
before -- back into a proliferating B-cell state.

A delay $\Delta t_2$ in the production
of plasma- and memory-cells with respect to the start of 
hypermutations was indeed observed experimentally 
(Jacob \et 1993; Pascual \et 1994b)
and is of the order of $48$ hours. This justifies the
separation of optimization and output production phase in the
model. In addition it has previously been shown that such a delay
is necessary in order to get an appropriate average quality
of produced output cells. This delay could be determined 
quantitatively using the experimental observation that the
number of produced output cells multiplies by a factor
of $6$ between day $6$ and day $12$ after immunization
(Han \et 1995b).
The steepness of the number of produced output cells versus time
is directly influenced by the duration of the optimization
phase. In this way one is led to a delay of the output
production phase of $48$ hours 
(Meyer-Hermann \et 2001),
which is in perfect
accordance with the above mentioned experimental observation.
Therefore, this value is assumed for the present model and
the resulting steepness of the output production will
be verified subsequently (see Sec.~\ref{reference}).
Nevertheless, it would be interesting to discuss possible
mechanisms which may generate the start of output production
dynamically.

Positively selected centrocytes differentiate into
output cells at a rate of $g_3=1/(7\,hr)$.
As we assumed recycling to exist, 
the proportion of differentiation into output
cells to differentiation into a re-proliferating cell state 
has to be chosen.
The recycling probability of positively selected
B-cells is determined by the only known experimental
evidence for recycling 
(Han \et 1995a).
A quantitative evaluation
(Meyer-Hermann \et 2001)
leads to a probability of $q=80\%$, i.e.~$4$ out of
$5$ positively selected B-cells do not differentiate into
plasma- or memory-cells but restart to proliferate and mutate,
and only one cell differentiates into an output-cell.

\subsubsection{The end of the GC reaction}

The whole GC reaction lasts about $21$ days 
(Jacob \et 1991; Liu \et 1991; Kelsoe, 1996).
No consumption of antigen fragments 
(Tew \& Mandel, 1979)
is postulated and no production of antibodies by
plasma cells is assumed
in order to stop the GC reaction. 
The model assumes that the main course
of the cell population dynamics is governed by the local
cell interactions and thus is independent of antigen consumption 
(Meyer-Hermann \et 2001).
As the amount of presented
antigen is reduced to $50\%$ during $30$ days
(Oprea \et 2000), 
i.e.~during the whole GC life time,
it is unlikely that the fast reduction of the total cell population 
after about $5$ days 
(Liu \et 1991; Hollowood \& Macartney 1992)
is due to antigen consumption. 

Note, that if plasma cells differentiate inside the GC
they may also produce antibodies there. 
These antibodies could bind the antigen fragments and in this way
reduce the probability of selection of centrocytes
if a large number of antibodies is present inside the GC.
This effect may become important at the end of the
GC reaction, when the number of plasma cells has already
increased. The final phase of the GC would be shortened
compared to the model results.
The resulting final state
of the GC reaction as it appears in this model will
be further discussed (see Sec.~\ref{volume}).


\subsection{Spatial distribution on a lattice}
\label{lattice}

One major task of the model is to find possible explanations
for the specific morphological appearance of GCs. Especially,
the dark and light zones are of interest, as it remains unclear
if they may be or why they are essential for affinity maturation.
A deeper understanding of these connections may be important
in view of malignant GCs as well
(Falini et al., 2000; Dunn-Walters et al., 2001). 

The spatial distribution
of the B-cells and FDCs is implemented in an equidistant lattice of fixed
volume. This constant volume is considered as reaction volume and thus is not
identical to the real GC volume. The latter is defined by the
volume occupied by all B-cells in the reaction volume and 
will consequently change in time. The resulting GC volume time course
is compared to experimental data from real GC reactions
in Sec.~\ref{volume}. Note, that in this prescription possible changes
of the GC dynamical behavior due to the interaction 
with naive B-cells in the mantle zone, for example, are neglected.

Each lattice point can contain one cell only. 
Signal molecules can accumulate at every lattice point. 
The cell movement is basically governed by diffusion (see Sec.~\ref{movements}).
The lattice constant is chosen to correspond to a typical B-cell diameter. 
Centroblasts have a radius of about $r_{\CB}\approx7.5\micron$ 
(Kroese \et 1987; Hostager \et 2000)
and centrocytes are substantially smaller 
(Liu \et 1994).
The average radius is of the order
of $5\micron$, and we are led to a lattice constant of $\Delta x=10\micron$.
However, this value solely has an influence on the cell mobility and therefore is 
not an independent parameter of the model but determines the meaning
of the diffusion constants (see Sec.~\ref{movements}).

\subsubsection{B-cell, FDC, and signal representation}
\label{represent}

In the following the specific treatment of the different cell types
on the lattice and the corresponding actions are explained in terms
of rules for cell behavior. 
In principle, every cell is characterized by its position
on the lattice, the state of differentiation, and the
type of encoded antibody in the shape space 
(see Sec.~\ref{shapespace}). 
Note that not all possible actions will occur in all phases of the GC
reaction (see Sec.~\ref{phases}). Note further that in general some
actions are forbidden to occur in the same time step. This is no relevant
restriction if the time steps are sufficiently small, so that in
most time steps no action occurs at all. 

\paragraph{Centroblasts}
can be in two different states: Originally, an 
{\it unsignaled} centroblast can proliferate,
mutate, and diffuse. When it meets a threshold concentration of the differentiation
signal it switches to the state {\it signaled} and can additionally 
differentiate into centrocytes (with rate $g_1$).\\
At each time step the concentration of signal molecules is
checked at the same lattice point. The state of the centroblast
is chosen correspondingly. In any case the B-cell remains a centroblast.
It then tries to proliferate (see Sec.~\ref{prolifspace}) and, 
provided that it is in the state {\it signaled}, 
tries to differentiate into centrocytes (with rate $g_1$).
Only one of the last two actions is allowed and both actions
are tried in a random sequence. 
If the centroblast proliferates both new centroblasts 
mutate (change the position in the shape space) with probability $m$.
If the centroblast did not differentiate to a centrocyte it might 
also diffuse with diffusion constant $D_{CB}$. 
A diffusion beyond the GC volume is forbidden.

\paragraph{Centrocytes}
emerge through differentiation of centroblasts only. They can be in
three different states: First they are {\it unselected}, may then get in
{\it contact} with FDCs, and finally may become positively {\it selected}.
\begin{itemize}
\item 
An unselected centrocyte dies with a rate $z$. 
A dead centrocyte is eliminated from the lattice.
\item
During its life time it
checks for an FDC on the neighbor lattice points. If there is one,
the switch to the state {\it contact}
is weighted with the affinity function Eq.~\ref{affinity}. 
A centrocyte which has tried in vain to get into contact with an FDC 
has to move by diffusion before it may try again (this introduces
an absolute time scale for the binding process which is basically
determined by the centrocyte diffusion constant $D_{\rm CC}$).
\item
A centrocyte in contact with an FDC dissociates from it and switches to the
state {\it selected} with rate $g_2$. It can neither move nor die.
\item
A positively selected centrocyte further differentiates with rate $g_3$
into an output cell with probability $1-q$ or
a centroblast with probability $q$. If the centrocyte did not differentiate
it might diffuse with diffusion constant $D_{CC}$.
\item
The diffusion of centrocytes beyond the GC volume is forbidden
independently of its state.
\end{itemize}

\paragraph{Output cells}
diffuse on the lattice with diffusion constant $D_{CC}$. They have
no restriction at the border of the GC volume, i.e.~if an output cell diffuses
beyond the GC volume it is eliminated from the lattice.
An eliminated output cell is memoried and contributes to the total
(i.e.~time integrated) output production.

\paragraph{FDCs}
are represented on the lattice by a center with
$2d$ arms of $30\micron$ length, where $d$ is the space dimension
of the lattice. These cells are assumed to be immobile.
This does not correspond to reality as at least the arms
of the FDCs are steadily looking for an interaction partner.
However, this is not an important restriction, since the movement 
of the FDCs would primarily change the interaction frequency 
of FDCs and centrocytes, which may be achieved by a faster 
centrocyte diffusion as well.

\paragraph{The differentiation signal}
is represented as a discretized density on the whole lattice.
One quantum of the signal is thought to contain exactly as many molecules
as are necessary to induce the differentiation process in one
centroblast. Those quanta move on the lattice according
to a classical diffusion equation (see Sec.~\ref{movements})
independently of the cell distribution on the lattice. 
In order to keep the quanta undivided, 
the movement of the remaining non-integer part of a quantum
is chosen randomly in each time step.
If an unsignaled centroblast is present on the same lattice point at some
time step, one of those quanta at this point is eliminated from the lattice
and the state of the centroblast is changed to {\it signaled}.
The signal molecules may diffuse beyond the GC volume.
In this case they are eliminated from the lattice.

\subsubsection{Initial cell distribution}

At the beginning of the GC reaction only FDCs and a few seeder
cells (centroblasts) are present in the follicle system.
The follicular network is completely filled during the
phase of monoclonal expansion 
(Liu \et 1991; Camacho \et 1998).
The centroblasts even need slightly more space than provided by the
FDC network. This is incorporated into the model by a random 
initial distribution of the FDC on a lattice area which is
restricted to about $70\%$ of the maximal GC volume.
However, for reasons of clarity and of comparability of the results,
all simulations presented in this article are based on the same
FDC distribution (see Sec.~\ref{stochastic}).
The number of seeder cells is chosen to be $3$
(Kroese \et 1987; Liu \et 1991; Jacob \et 1991; K\"uppers \et 1993).
They are randomly distributed on the FDC network.

\subsubsection{Spatial aspects of proliferation}
\label{prolifspace}

The proliferation process itself is not only dependent
on the proliferation rate but also on the environment
of the dividing centroblast. Principally, a centroblast
can only proliferate if there is a free lattice point
in its neighborhood. On the other hand, this scenario 
is not realistic, as the centroblasts may deform and
the dividing process does conserve the total volume of
both new cells. A strong exclusion rule that forbids
proliferation if all neighbor points are occupied would
therefore strongly decrease the number of proliferations
compared to a realistic or exponential increase
of the total cell number. The solution we propose is to allow
the proliferation if there is a free lattice point
within a sphere around the dividing cell. The centroblasts
in the environment of the dividing cell are rearranged
in such a way that the proliferation becomes possible.
In this picture a dividing cell is able to push away a 
maximal number of cells which corresponds to the
radius $R_{\rm P}$ of this theoretical sphere.
Within the sphere the centroblasts
expand exponentially. When the centroblasts
expand beyond our theoretical sphere a nucleus of centroblasts,
which are no longer able to proliferate until
the surrounding cells are spread through diffusion, emerges,
and the exponential growth is slowed down.
Therefore, the resulting total number of centroblasts is directly
changed in dependence on the sphere radius.
As the number of centroblasts after $3$ days of the
GC reaction is known experimentally, the radius is
chosen correspondingly, where a value of $R_{\rm P}=60\micron$
appears to be reasonable.
This spatial proliferation model has realistic features which
are sufficient for our purpose as we are not primarily interested
in the details of the proliferation process. Nevertheless, a more
detailed mechanistic model may lead to new dynamical
insights (Beyer \& Meyer-Hermann, 2002).

\subsubsection{Movement of cells and molecules}
\label{movements}

Centroblasts, centrocytes, and signal molecules essentially move
by diffusion. 
This is realized on the lattice by a probability of moving to a free
neighbor point on the lattice which can be calculated
using the classical diffusion equation
\be{diffusion}
\frac{\partial c}{\partial t}
\;=\;
D\,\sum_{i=1}^d \frac{\partial^2 c}{\partial x_i^2}
\quad,
\ee
where $c$ is the population of the diffusing cell type,
$d$ denotes the dimension of the lattice, $t$ the time, and
$x_i$ the spatial coordinates. 
This equation is translated to a discrete equation with
lattice constant $\Delta x = \Delta x_i$ (the lattice is
equidistant) and time step $\Delta t$
\be{diffdiscrete}
\frac{\Delta c}{\Delta t}
\;=\;
D\,\sum_{i=1}^d \frac{\Delta^2 c}{\Delta x^2}
\;=\;
D\,\frac{-N}{\Delta x^2}
\quad,
\ee
where $N$ is the number of free neighbor points that can adopt
values between $0$ and $2d$. Using this one may define a
statistical probability $p_{\rm diff}$ for one cell to move
to a free neighbor point:
\be{diffprobable}
p_{\rm diff} \;=\;
D\,N\,\frac{\Delta t}{\Delta x^2}
\quad.
\ee
One should be aware that this probability is based on the assumption that
each lattice point can be occupied by one cell only. Therefore,
the diffusion of signal molecules, which do not underly this
restriction, has to be treated according
to the more general first part of \gl{diffdiscrete}.

It is worth emphasizing that the assumption to allow only one
centroblast or centrocyte at each lattice point, independently of the 
volume occupied by these cells,
does not imply the neglection of the differing cell diameters.
The cell volume primarily has an impact on the mobility of the cells. 
Therefore, the cell diameters are incorporated into the model
by using different diffusion constants $D_{\CB}$ and $D_{\CC}$ for
centroblasts and centrocytes, respectively. Quantitatively, the
relation between both parameters is determined by the relation
between the diameters: Considering the cells as spheres of a specific
radius $r$ moving in a medium of viscosity $\eta$ at temperature
$T$ under influence of friction according to Stokes, the diffusion
constant becomes
\be{diffusconstant}
D\;=\; \frac{kT}{6\pi\eta r}
\quad,
\ee
where $k=1.38066 10^{-23} J/K$ is the Boltzmann constant. At
room temperature $T=293\,K$, with a viscosity for blood
of $\eta=0.02 Js/m^3$ 
(Skalak \& Chien, 1987),
and with a cell radius
of $2.5\micron$ and $7.5\micron$ 
for centrocytes and centroblasts,
respectively, (Kroese \et 1987)
(see also 
Liu \et 1994
for the ratio of both radii)
one finds
\be{diffus2constant}
D_\CB \;\approx\; 5\,\frac{\micron^2}{hr}
\quad ; \quad
D_\CC \;\approx\; 15\,\frac{\micron^2}{hr}
\quad.
\ee
Surely, the absolute values of these diffusion constants have been 
estimated very roughly. Especially, the lymphocytes do not diffuse 
in blood but in lymphoid organs.
Additionally, the movement of the FDCs effectively increases the
observed diffusion constant in the model.
Finally, the assumption of a friction according to Stokes may
be questioned.
Nevertheless, the assumption that the diffusion constants depend
on the cell radii is reasonable, and the ratio of the radii is relatively
sure. Therefore, the main uncertainty concerns the absolute values of the
diffusion constants, not their ratio (see Sec.~\ref{robdiffuse}).

The diffusion of signal molecules is assumed to be 
considerably faster than the diffusion of the cells,
with the diffusion constant being of the order of (Murray, 1993)
%
\be{diffussignal}
D_\Sig \;\approx\; 200\,\frac{\micron^2}{hr}
\quad.
\ee

\subsubsection{A two-dimensional projection}
\label{2Dprojection}

In order to compare the model results with pictures of slices
containing GCs, the model is realized on a two-dimensional (2D)
lattice. One should be aware that a three-dimensional (3D)
model may open new pathways for cell movements, so
that a more realistic three-dimensional model may lead to
different dynamical properties in some aspects. Nevertheless, 
we expect the basic properties to remain untouched from this
projection on a two-dimensional lattice. The comparison
of the results in a 2D and a 3D model is the task of a
separate work.

It is important to project the physiological parameters on the 2D
lattice. This concerns all parameters which have an impact on the
spatial distribution of the cells in the GC, i.e.~the 
proliferation rate, which controls the creation of
new cells at specific positions, the number and the geometry
of the FDCs, and the total volume of the GC. 
The diffusion constants are defined
for an arbitrary lattice dimension (see above) and the
diameters of the spheric cells remain unchanged in a two
dimensional plane. 

\paragraph{Proliferation rate}
The proliferation rate in a 3D-GC of
$1/(6hr)$ corresponds to $1/(9hr)$ in a 2D lattice, as only
$4$ of $6$ new cells emerging from a proliferation process
will appear in the 2D plane. Therefore, the final number
of B-cells after $3$ days of monoclonal expansion will not
be $12288$ as in an idealized 3D model but $768$. 

\paragraph{GC-volume}
The maximal number of cells determines the total size of the 
2D lattice. According to experimental observations 
(Liu \et 1991; Camacho \et 1998)
the FDC network, i.e.~the volume
of the GC is completely filled with centroblasts after $3$ days
of proliferation. 
Starting from a typical 
GC volume three days after immunization of $0.05\,mm^3$,
we end up with a 
GC radius of $232\micron$ or $47$ lattice points per dimension.
Based on this rough estimate, the number of  lattice points
per dimension is fixed to $45$ corresponding to a radius
of $220\micron$. This still allows a certain cell movement even
if the cell number in the GC is at its maximum.

\paragraph{FDCs}
\label{FDC2D}
The number of arms of the FDCs is reduced from $6$ (3D) to
$4$ in the 2D model. The length of the FDC arms remains unchanged. 
It is worth emphasizing that the FDC arm length and the number
of FDCs together determine the available surface for interaction
with centrocytes, and in this way influence the interaction frequency of 
centrocytes and FDCs. 
The total number of FDCs is chosen in such a way that the centrocytes have
enough space to move and at the same time have enough possibilities
to interact with FDCs, i.e.~the number of binding
sites on the FDCs should be of the order of the maximal number 
of centrocytes.
As centrocytes differentiate from centroblasts
and the centroblast differentiation rate is of the same order as the 
centrocyte apoptosis rate, the maximal number of centrocytes
will lie roughly in the same range as the maximal number of centroblasts.
For an FDC arm length of
$30\micron$ this leads to about $20$ FDCs 
in a 2D-GC with a radius of $220\micron$.


\section{Results}
\label{results}

The previous section was devoted to the physiological
motivation of model assumptions and to the determination of
the model parameters by experimental data. In the present
section the results of the numerical realization of this
stochastic model will be presented. 
As in some cases the experimental constraints for the
physiological quantities were not strong enough, there
still remains a certain freedom of choice for some of them.
Therefore, these quantities will have to be optimized
iteratively with respect to the outcome
within their physiological range.
The parameters in Tab.~\ref{parameter} turned out to be
an optimal choice with respect to different aspects of
the GC reaction, which will be described below.

However, the presentation of the results will not start
from these optimal values, since the necessity of some model
assumptions is better illustrated by going through the
different stages of the model development. Therefore, the
discussion starts with the consideration of different scenarios
for the establishment of the dark zone during GC reactions
(see Sec.~\ref{origin}).
Based on the results of this subsection 
a reference GC is defined (see Sec.~\ref{reference}).
Afterwards, it is discussed how the duration of the dark zone is 
determined (see Sec.~\ref{duration}), and 
the influence of B-cell recycling and the dark zone
on the optimization process, i.e.~on affinity maturation,
is investigated (see Sec.~\ref{quality}).
Finally, we discuss the evolution of the total GC volume
(see Sec.~\ref{volume}).

\subsection{Requirements for dark zones}
\label{origin}

\paragraph{FDC inhomogeneity}
An important task of this work is to investigate necessary and sufficient
conditions for the emergence of a dark zone during GC reactions.
The seed of a dark zone appears to
be an inhomogeneity of the FDC distribution on the GC volume.
During the first phase of the reaction
the initial centroblasts proliferate 
beside and/or inside the FDC network. However, at least some
of the centroblasts proliferate outside the FDC network or
expand beyond it. It is exactly these centroblasts that 
give rise to the development of the dark zone. As in
the model the maximal GC volume (i.e.~the lattice volume)
is fixed a priori, this dynamical behavior
is simulated by a statistical distribution of the FDCs on a restricted area
of the GC which occupies about $70\%$ of the total GC volume.
Without such a restricted distribution of the FDCs 
the distribution of centroblasts
and centrocytes on the GC necessarily becomes homogeneous. 
Accordingly, the expansion of centroblasts beyond or beside the
FDC network is a necessary condition for the development
of dark zones, and is independent of the mechanism 
which is assumed for the initiation of centroblast differentiation into
centrocytes. In this perspective, centroblasts,
which primarily proliferate outside the FDC network
(Camacho et al., 1998), would fascilitate the emergence of a dark zone.
In the following we start from the more modest point of view, that
only a part of the centroblasts is present outside the FDC network
after $3$ days of proliferation.

\paragraph{Centroblast specific differentiation}
As will be shown in the following,
the centroblast differentiation process is a key element
in the development of the dark zone.
In a first step we do not investigate any interaction of FDCs with
centroblasts or any signal molecules that
trigger the differentiation of centroblasts to centrocytes.
One may imagine for example the centroblasts to have an internal clock
which lets them differentiate after some more or less fixed time.
Alternatively, a fixed number of proliferation rounds for each centroblast
may be supposed. In both scenarios
the differentiation should merely start after three days of monoclonal
centroblast expansion, as centrocytes are not present in GCs before
this time. This type of hypothesis was tested in the present model and 
the expected result was found: 
No dark zone appears at all at any moment of the reaction.
This can easily be understood, as the differentiation process is
started everywhere in the GC volume at the same time. 
The differentiating
centroblasts are distributed homogeneously, especially 
in an area which has the potential of becoming the dark zone, i.e.~an
area without FDCs. This process
ends up with a homogeneous distribution of centroblasts and centrocytes
over the whole GC volume.

\begin{figure}[ht]
\vspace*{-20mm}
\begin{center}
\includegraphics[height=6.9cm]{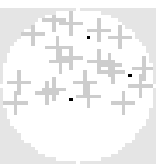}
\includegraphics[height=6.9cm]{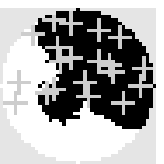}\\[0.3ex]
\includegraphics[height=6.9cm]{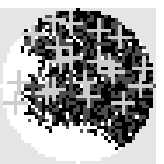}
\includegraphics[height=6.9cm]{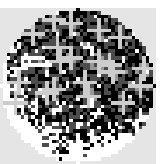}\\[0.3ex]
\includegraphics[height=6.9cm]{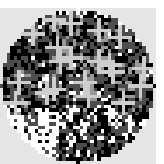}
\includegraphics[height=6.9cm]{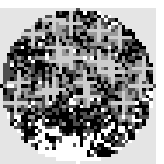}\\
\vspace*{-21.0cm}
\begin{minipage}[t]{14cm}
\hspace*{2mm} day $0$
\hspace*{57mm} day $3$\\[64mm]
\hspace*{2mm} day $4$
\hspace*{57mm} day $8$\\[64mm]
\hspace*{2mm} day $12$
\hspace*{55mm} day $21$
\end{minipage}
\vspace*{68mm}
\end{center}
\vspace*{-8mm}
\caption[]{\sf 
If the centroblast differentiation is independent of
the FDCs (especially without differentiation signals)
no dark zone appears during the GC reaction.
This remains true even with a centrocyte attractor in the FDC
network. Here, the centrocyte-FDC interaction time 
is $g_2^{-1}=10\,hr$, and the
centroblast differentiation rate is $g_1=1/(8.7\,hr)$. 
All other parameters are like in Tab.~\ref{parameter}.
The initial distribution and the stages of the reaction after 
$3,4,8,12,21$ days are shown. 
The represented cell types are (from dark to light):
proliferating cells, centrocytes (in all stages of the selection process),
output cells, FDCs, and empty space.}
\label{0024c}
\end{figure}

As a consequence, in the framework of the above scenarios the appearance 
of dark zones is necessarily coupled to an additional force, driving the separation 
of centroblasts and centrocytes. One may investigate mechanical
forces in this context, since both cell types differ in size. This has
an impact on their diffusion constant. The radii differ
by a factor of at least $2$ 
(Liu \et 1994),
so that the distance the cells may
overcome by random walk differs by a factor of about $1.4$. 
Apparently, there exists
a slight difference in the mobility of both cell types, but without
definite direction. The separation of both cell types may
be achieved either by a local attractor for one cell type 
or by a homogeneous force field.
The latter possibility corresponds to the problem of sorting large and small
balls in the field of gravity and is presently under investigation 
(Beyer \& Meyer-Hermann, 2002).
This possibility cannot be excluded here, even if 
the origin of such a force field remains unclear from the physiological
point of view.
At first sight, the existence of a local attractor seems to be a promising assumption
as the FDCs may bind only centrocytes but no
centroblasts. We tested how
long this centrocyte-FDC binding process has to endure in order to get
a realistic separation of centrocytes and centroblasts. Unfortunately,
no dark zone appears at all (see Fig.~\ref{0024c}),
even using extremely long centrocyte-FDC
interaction times of about $100\,hr$, which is completely unreasonable as
in experiment the binding time is estimated to be of the order
of $2~hr$ 
(Lindhout \et 1995; van Eijk \& de Groot, 1999).
The reason is that the centroblast differentiation occurs
independently of the position in the GC, i.e.~homogeneously distributed.
Even if there was an increased amount of
centroblasts in the area without FDCs, this tendency to a 
dark zone would be destroyed in the same moment due to
the centroblast differentiation process, which takes place
everywhere in the dark zone. Therefore, in this scenario
one cannot expect to get stable dark zones. They could at the best
appear as statistical fluctuations of the centroblast population.

\begin{figure}[ht]
\begin{center}
\includegraphics[height=5.3cm]{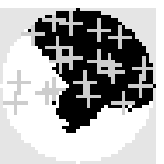}
\includegraphics[height=5.3cm]{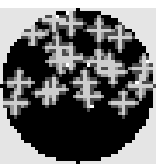}
\includegraphics[height=5.3cm]{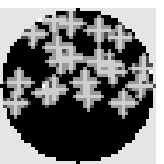}\\
\vspace*{-5.5cm}
\begin{minipage}[t]{14cm}
\hspace*{-10mm} day $3$
\hspace*{41mm} day $8$
\hspace*{42mm} day $21$
\end{minipage}
\vspace*{47mm}
\end{center}
\vspace*{-3mm}
\caption[]{\sf Using the concept of centroblast differentiation by 
local interaction with FDCs, the dark zone is established very
early and is stable until the end of the GC reaction.
The parameters are like in Tab.~\ref{parameter}, except
that the centroblast differentiation occurs in the moment
of contact with FDCs (realistic differentiation rates further
enhance the fraction of centroblasts) and that no
differentiation signal exists.
The stages of the reaction after 
$3,8,21$ days are shown. 
The represented cell types are (from dark to light):
proliferating cells, centrocytes (in all stages of the selection process),
output cells, FDCs, and empty space.}
\label{0025c}
\end{figure}

\paragraph{Local centroblast-FDC interaction}
Now let us turn to a third scenario, where the FDCs initiate the
centroblast differentiation by a local interaction.
In this case only those centroblasts which
are placed at the surface of an already established dark zone
would differentiate into centrocytes and the nucleus of the 
dark zone would remain unaffected by the differentiation at the
surface. Therefore, a stable establishment of a dark zone
is possible in this scenario. This gives rise to two important
questions: How long does a dark zone exist? How does the
GC reaction stop? Both questions are answered very clearly
by the model and the answers are stable against variations
of the model parameters: The dark zone develops early and
remains present until the end of the reaction. The total cell population
in the GC does not decline but approaches a constant. The FDCs are
surrounded by centrocytes which are in selection process, so
that spots of FDCs with centrocytes appear on a background
of centroblasts (see Fig.~\ref{0025c}).
The fraction of centrocytes in the FDC network is governed 
by the selection rates and the FDC density: For large centroblast-
and large (selected) centrocyte-differentiation rates and for large
FDC densities the fraction of centrocytes can be increased.
But the qualitative appearance of the GC does not change.
This spot-like GC is not in agreement with experimentally
observed GCs. Those have no spots and after a while
one finds more centrocytes than centroblasts. 
In addition, the GC reaction
declines after 3 weeks. In the model a reduction of the total
cell population can only be achieved by an additional assumption
concerning the proliferation rate. Taking into account that
the dark zone has been observed to disappear after $8$ days
(Camacho \et 1998)
and that the total GC volume already decreases at day $5$
(Liu \et 1991; Hollowood \& Macartney 1992),
the proliferation rate would have to be substantially diminished
in this early phase of the GC reaction.
It is very unlikely that this is
really the case, since the proliferation rate has been frequently measured  
to be of the order of $1/(6hr)$, and this value was found at
different moments of the GC reaction
(Hanna, 1964; Zhang \et 1988; Liu \et 1991). 

\paragraph{Non-local centroblast-FDC interaction}
As has been shown, centroblast-specific differentiation scenarios,
even being combined with an attractor
which incorporates an inhomogeneity into the GC volume, do not
lead to the expected appearance of dark zones.
A centroblast differentiation process which is governed by a 
local interaction with FDCs
leads to conceptual problems concerning the 
centroblast proliferation rate. 

Therefore, in a fourth scenario,
a nonlocal interaction of centroblasts and FDCs 
is investigated in order to find
the experimentally observed morphology. Such a nonlocal interaction
may be mediated by a diffusing signal molecule which is produced
by the FDCs during the GC reaction. Each time when a centroblasts
meets a threshold density of signal molecules it starts to differentiate 
into centrocytes at a certain rate. 
The signal molecules that initiated the differentiation process are
not eliminated from the lattice, because they are not consumed by the
centroblasts.
The model then predicts a short appearance
of a dark zone which is depleted within hours. The fast depletion 
is due to the large diffusion constant for the signal molecule in comparison
to the diffusion constant of centroblasts. Therefore, the signal
molecules become homogeneously distributed on the whole GC volume,
especially in the dark zone, within a short time. A homogeneous
distribution of the differentiation signal is undistinguishable from
the first scenario {\it Centroblast specific differentiation}:
All centroblasts differentiate into centrocytes, independently of
their position.
Consequently, the dark zone is destroyed in the same way
as in the scenario discussed before. In the present scenario
dark zones cannot survive until the diffusion constant of the 
signal molecules is reduced to an unreasonable small value.

This leads us to a fifth scenario which builds upon the previous one
by incorporating the consumption of signal molecules. It is rather plausible
that the signal molecules have to bind to the centroblasts
in order to initiate the differentiation process. 
Then, bound molecules are not available for a second initiation
of centroblast differentiation as they have been before. We investigated
this scenario in the model: As the FDC network is the source
of signal molecules, only the centroblasts
at the surface of the dark zone go into interaction with
the differentiation signal and consume it. 
Therefore, the nucleus of the dark zone
is widely protected against contact with signal molecules and can
survive for significantly longer periods. 

\begin{figure}[ht]
\vspace*{-5mm}
\begin{center}
\includegraphics[height=6.9cm]{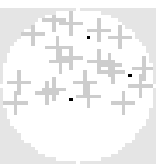}
\includegraphics[height=6.9cm]{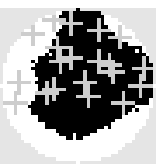}\\[0.3ex]
\includegraphics[height=6.9cm]{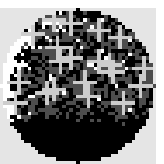}
\includegraphics[height=6.9cm]{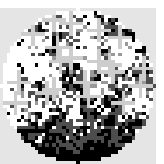}\\[0.3ex]
\includegraphics[height=6.9cm]{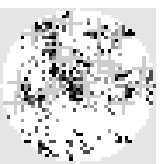}
\includegraphics[height=6.9cm]{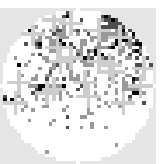}\\
\vspace*{-21.0cm}
\begin{minipage}[t]{14cm}
\hspace*{2mm} day $0$
\hspace*{57mm} day $3$\\[64mm]
\hspace*{2mm} day $4$
\hspace*{57mm} day $8$\\[64mm]
\hspace*{2mm} day $12$
\hspace*{55mm} day $21$
\end{minipage}
\vspace*{68mm}
\end{center}
\vspace*{-6mm}
\caption[]{\sf Development of the dark zone during a GC reaction: The
initial distribution and the stages of the reaction after $3,4,8,12,21$ days.
The represented cell types are (from dark to light):
proliferating cells, centrocytes (in all stages of the selection process),
output cells, FDCs, and empty space.
The used parameters are given in Tab.~\ref{parameter}.}
\label{0x29axx}
\end{figure}

A representative simulation result is shown in Fig.~\ref{0x29axx}: 
A large amount
of proliferating centroblasts develops from $3$
seeder cells. Those centroblasts which are placed beside
the FDC network give rise to the dark zone (in the lower
hemisphere). The dark zone is stable for a certain time and 
{\it evaporates} cells at its surface by differentiation into
centrocytes in dependence on the diffusing
signal molecules. 
The centrocytes generated at the interphase of the dark and the light
zone tend to diffuse into the FDC network.
The resulting life time of the dark zone directly depends 
on the production rate of signal molecules.
When the dark zone has disappeared, centroblasts
and centrocytes are distributed more or less
homogeneously on the GC volume. During the remaining
time the GC reaction declines.

Summarizing, we find two conditions which are together sufficient for the
appearance of dark zones:
\begin{enumerate}
\item A monoclonal expansion of centroblasts beyond the FDC network.
\item A nonlocal interaction between FDCs and centroblasts, which
may be realized with a differentiation signal molecule which is 
produced by FDCs and bound by centroblasts.
\end{enumerate}
As has already been argued at the beginning of this section (see
the subsection on {\it FDC inhomogeneity}) condition $1$ is
also a necessary condition for the development of the dark zone. 
However, concerning the non-locality of the centroblast-FDC interaction
the situation is less clear and the argument for the necessity of this
condition is a negative one: All alternative scenarios for the
centroblast differentiation process lead to conceptual problems.
As a consequence, non-locality cannot be deduced to be
an imperative feature of the centroblast-FDC interaction.
Nevertheless, the reader may agree that the presented
scenario is an attractive and plausible way to explain
the development of dark zones in GC reactions as it has
been observed in experiment.


\subsection{The reference germinal center}
\label{reference}

After having determined features of GC reactions that are
sufficient for the development of dark zones, a reference GC can be
defined.
As we are dealing with a stochastic
model we have to analyse ensembles of simulations with
the same set of parameters but different generators 
of random numbers. Therefore, the results will always be
given with an uncertainty corresponding to one standard deviation.
The parameters in Tab.~\ref{parameter} that
were mostly determined in Sec.~\ref{modell},
are used as a reference system.
Nevertheless,
some values had to be determined by an iterative
optimization procedure depending on the outcome of the model.
All simulations were done with time steps of
$0.004~hr$. A dependence of the results on the time resolution
has not been observed in this regime.
In each time step all lattice points are actualized in a
random sequence.
In the following, the general character of the reference
GC will be shortly described. Special aspects
will follow in the subsequent sections.

\begin{figure}[ht]
\begin{center}
\includegraphics[height=6.9cm]{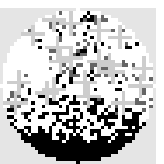}
\includegraphics[height=6.9cm]{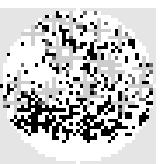}\\
\vspace*{-7.15cm}
\begin{minipage}[t]{14cm}
\hspace*{0mm} centroblasts
\hspace*{46.2mm} centrocytes
\end{minipage}
\vspace*{62mm}
\end{center}
\vspace*{-3mm}
\caption[]{\sf The same GC reaction as in Fig.~\ref{0x29axx}
is shown at day $6$ for
proliferating cells (left) and centrocytes (right) separately.
The recycled proliferating B-cells (grey, left figure) do not
reenter the dark zone, where only non-recycled centroblasts
(black, left figure) are found. The unselected centrocytes 
(black, right figure)
are selected (grey, right figure) in the narrow neighborhood of the FDCs
(light grey).}
\label{cbcc0x29a}
\end{figure}

The simulation shows the three phases that were postulated:
In the first phase the three seeder cells proliferate, so that
the B-cell population grows exponentially. After $3$ days
the increase is slowed down due to spatial restrictions and
due to the start of the centroblast differentiation process,
which is governed by the signal molecules produced by the
FDCs. The dark zone develops as shown in the previous section
and vanishes at day $8.2\pm 0.4$ (see Sec.~\ref{duration}).
The selection of centrocytes takes place in the environment
of the FDCs (see Fig.~\ref{cbcc0x29a}).
At day $5$ the production of output cells is started.
It has been observed in experiment that 
between day $6$ and $12$ of the reaction the number of output
cells multiplies $6$-fold (Han \et 1995b). 
In the model this number is $5.9\pm1.2$.
The whole GC population, i.e.~proliferating B-cells and centrocytes,
decreases during this last phase
and reaches a final number of $49\pm29$ proliferating B-cells
after $21$ days of the reaction. This number is influenced
by the variation (within the physiological constraints)
of the speed of the selection
process, especially by changes of the centroblast differentiation
rate (see also Sec.~\ref{volume}).


\subsection{Duration of the dark zone}
\label{duration}

Experimentally, the dark zone has been observed in a lot of different
GCs. It has been observed that the dark zone is already present $4$ or $5$ days
after immunization and vanishes approximately at day $8$
(Camacho \et 1998).
In the latter experiment no dark zone has been
observed at later times. 
Nevertheless, it remains unclear if this observation
is independent of the system under consideration and of the
experimental conditions. GCs with long lasting dark zones
may exist. On the other hand, one may suspect that the
complete absence of a dark zone is a sign for a malignant
immune reaction 
(Loeffler \& Stein, 2001).

\begin{figure}[ht]
\begin{center}
\includegraphics[height=11cm]{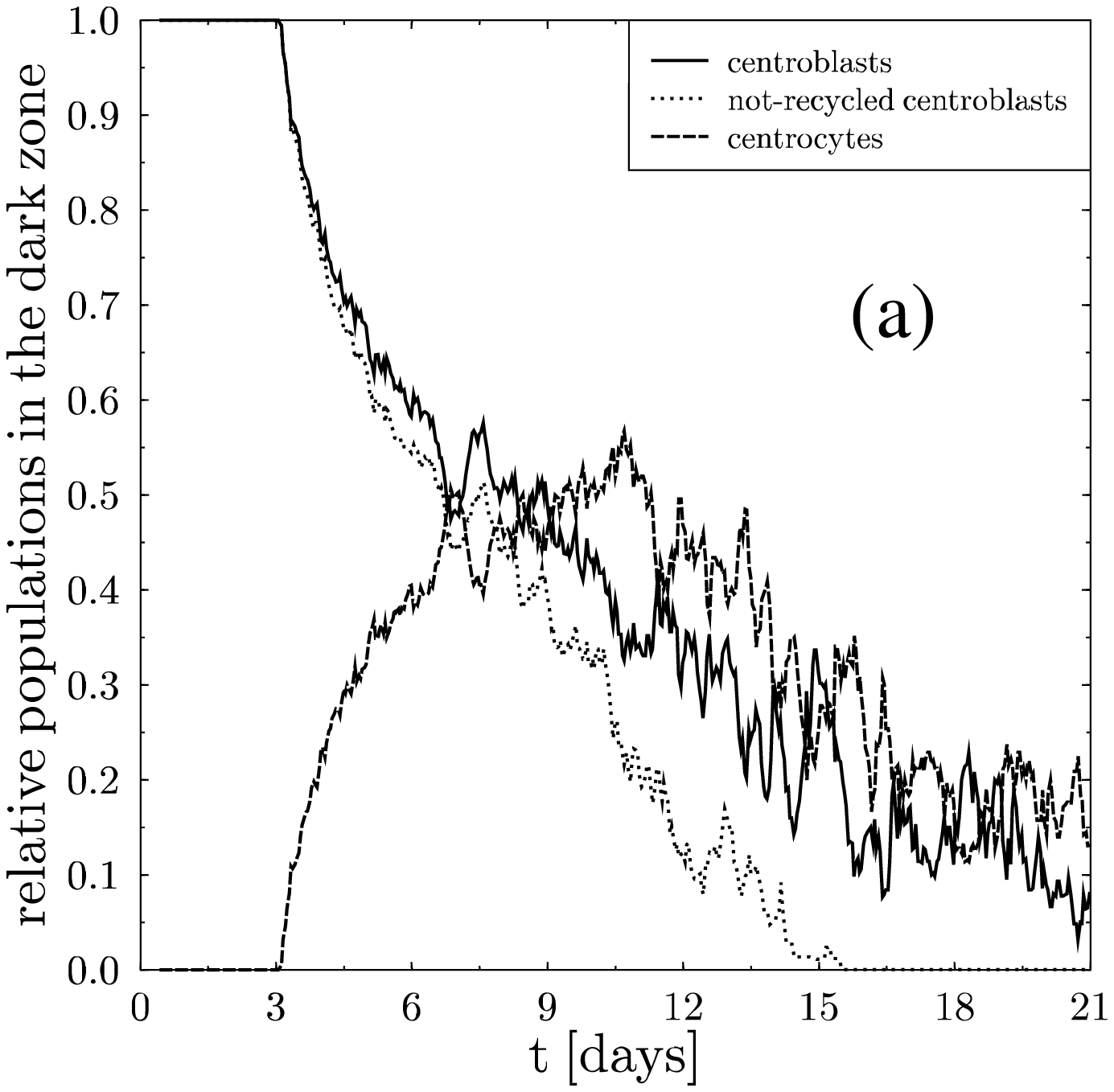}\\
\includegraphics[height=8cm]{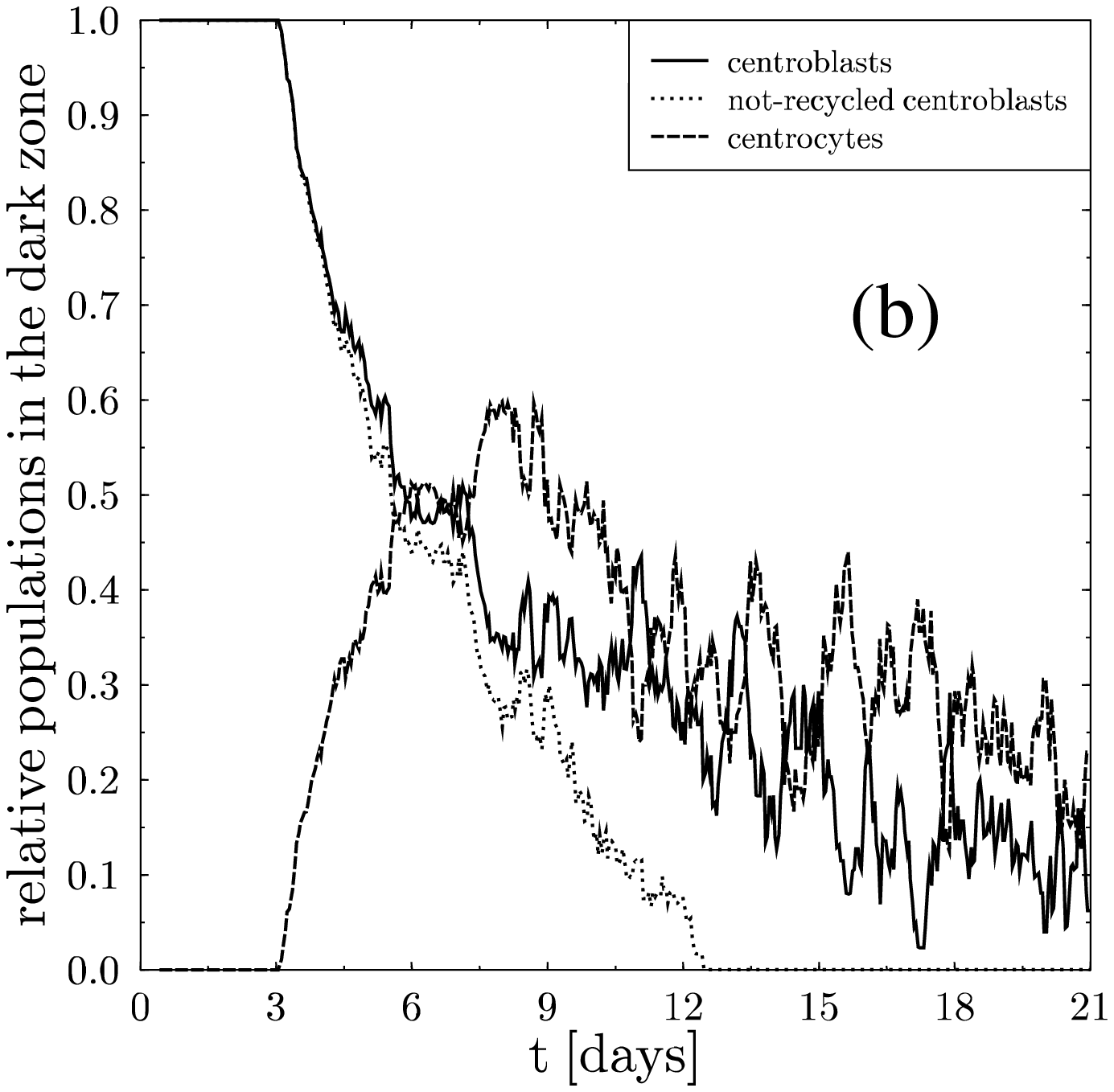}
\includegraphics[height=8cm]{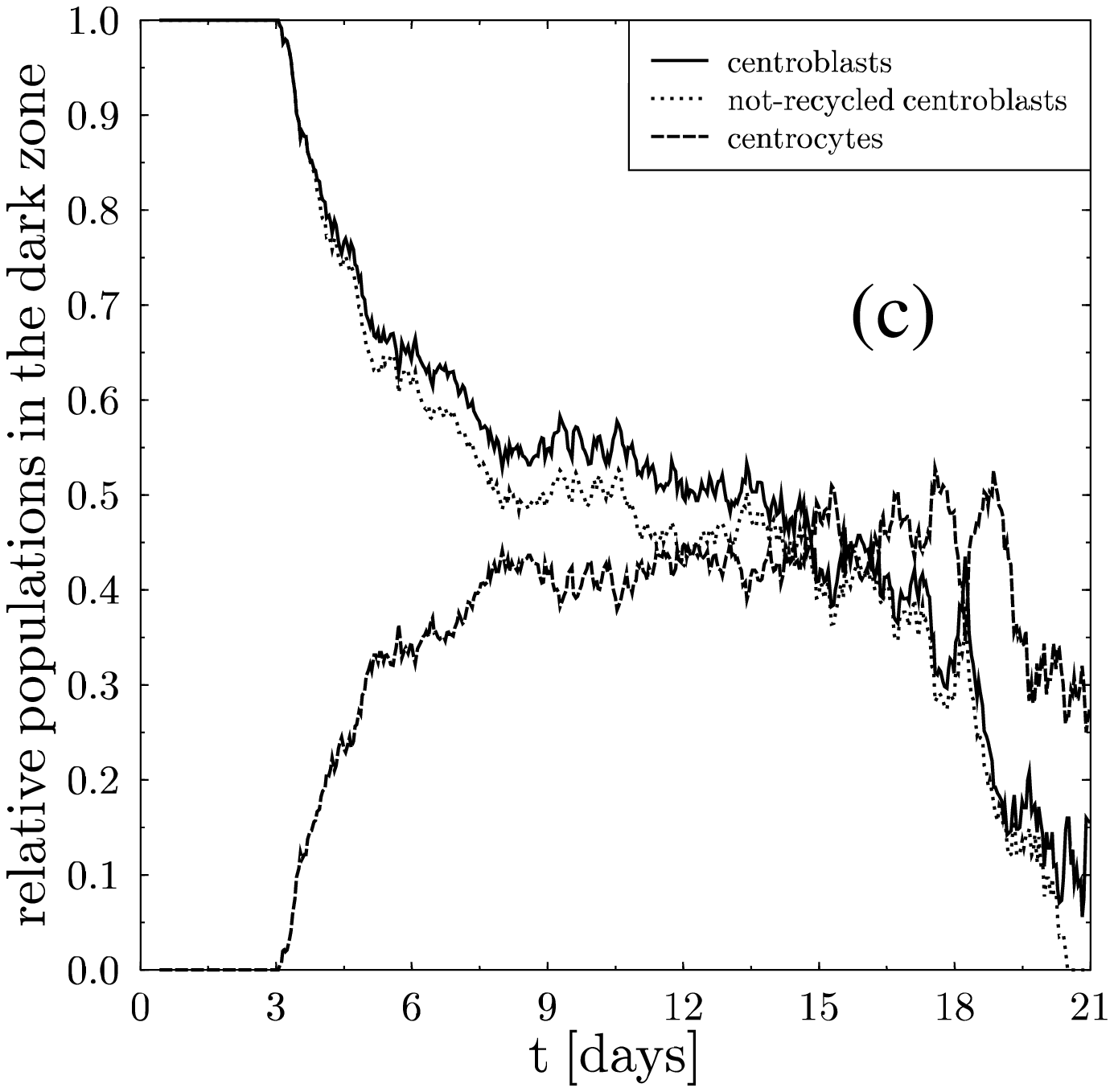}
\end{center}
\vspace*{-5mm}
\caption[]{\sf Representative time courses of the relative centroblast 
and centrocyte populations in the lower hemisphere of the GC
for different durations of the dark zone.
In (a) the parameters in Tab.~\ref{parameter} were used.
The production rate of signal molecules has been
changed to $s=6/hr$ (and the apoptosis rate to $z=1/8\,hr$) in (b) 
and to $s=12/hr$ in (c).
The dark zone breaks down after $8.3$ days (a), 
$6.9$ days (b), $14.9$ days (c).
Recycled B-cells have practically no impact on the
dark zone population.}
\label{zone_da}
\end{figure}

In the following we will analyse what determines the duration
of the intermediate appearance of dark zones. It is obvious
that the amount of signal molecules in the GC will influence
the speed of dark zone reduction by centroblast differentiation.
In order to quantify this statement, the dark zone is said
to have vanished if the number of centrocytes in the lower
hemisphere dominates the number of centroblasts. 
In Fig.~\ref{zone_da} this corresponds to the
crossing of the relative centroblast and centrocyte populations.
The dark zone vanishes at day $8.2\pm0.4$.
A variation of the signal molecule production rate (or equivalently of
the threshold concentration that is necessary to induce centroblast
differentiation) shows
that the life time of the dark zone indeed depends on this
parameter (see Fig.~\ref{zone_da} (b) and (c)).
Principally, all durations of dark zones may be produced by
a corresponding choice of the signal production rate. Therefore, the
concept of non-local centroblast-FDC interaction
allows intermediate dark zones as well as dark zones which 
remain stable during the whole GC reaction.

The duration of the dark zone is also influenced by other
physiological magnitudes.
One finds an interesting interplay of
the signal production rate, the rate of centroblast differentiation,
and the rate of centrocyte apoptosis. A slower centroblast
differentiation prolongs the total GC reaction, i.e.~more
B-cells remain after $21$ days.  Naturally, this
especially applies to the duration of the dark zone. 

The apoptosis rate has a slightly more complex effect.
This rate is the inverse centrocyte life time which,
at the same time, determines how much time a centrocyte is given
to be rescued from apoptosis by interaction with FDCs.
A smaller apoptosis rate of $1/(10hr)$ would still be in accordance
with experiment 
(Liu \et 1994).
We observe that the prolonged life time of
centrocytes has two basic effects: the duration of the dark zone 
is shortened and the total cell population is increased, so that the total
GC reaction is prolonged. The latter effect is rather intuitive,
as the death rate of centrocytes is reduced.
At the same time this implies that, in average, the number
of centrocytes is increased, especially, when compared to
the number of centroblasts. Since the moment of break-down
of the dark zone is defined by the relative numbers of
centroblasts and centrocytes, the duration of the dark
zone is shortened in this way.

This section is concluded with the remark, that the presented
non-local concept for centroblast differentiation is compatible
with almost all intermediate or non-intermediate appearances
of GC dark zones. The moment of dark zone destruction is
basically determined by 
the (experimentally unknown) signal production rate, 
the centroblast differentiation rate,
and the life time of centrocytes.

\subsection{Optimal optimization}
\label{quality}

Until here, the origin of the dark zone has been enlightened and
its stability during the GC reaction has been analysed. 
Now, the interest turns towards its function in the GC reaction,
especially for the affinity maturation process. 
At first, the affinity maturation process will be described
from the point of view of the model. Then the interplay
of recycling and dark zone, and its importance for affinity maturation 
will be pointed out.
Finally, the influence of the dark zone on the achieved
affinity to the antigen will be discussed.

\subsubsection{Affinity maturation}

\begin{figure}[ht]
\begin{center}
\includegraphics[height=11cm]{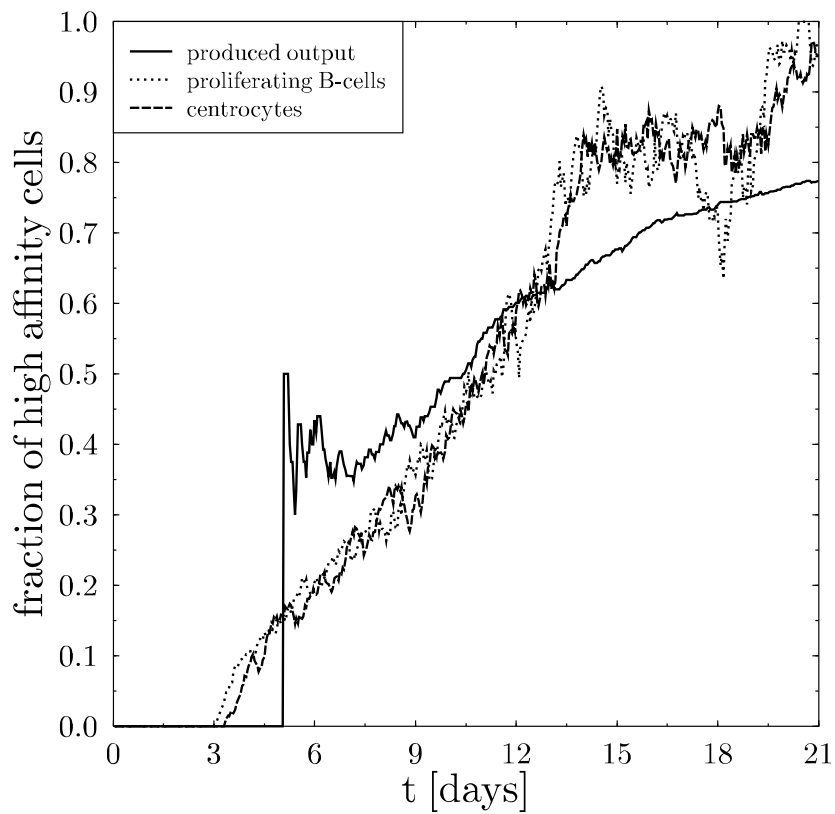}
\end{center}
\vspace*{-5mm}
\caption[]{\sf The time course of the fraction of high
affinity cells (cells which bind with a probability of at
least $30\%$) in the GC reaction is shown for centroblasts,
centrocytes, and for the sum of all output 
cells produced during the whole GC reaction.
The parameters in Tab.~\ref{parameter} were used.}
\label{0x29aaff}
\end{figure}

Among the major tasks of the GC reaction is the optimization of 
the antibody type with respect to the presented antigen
and the production of plasma cells which secrete these 
optimized antibodies. Therefore, it is important to verify
that a model which aims to help understand the origin and the
function of the GC morphology also describes the affinity
maturation process. 
B-cells are considered high affinity cells if they have
at least a $30\%$ probability to bind the antigen 
in the sense of Eq.~(\ref{affinity}), i.e.~if $a(\phi,\phi^*)>0.3$.
The time course of the thus defined fraction
of high affinity cells is depicted in Fig.~\ref{0x29aaff}.
After about $9.4\pm1.4$ days 
the high affinity cells dominate
in the GC, which compares to about $8$ days 
observed in experiment (Jacob \et 1993).

Figs.~\ref{0x29aaff} and \ref{cbandaff} show that the 
kinetics of the fraction of high affinity B-cells can be
grouped into four phases:
During the phase of monoclonal expansion no high affinity
B-cells exist at all. As no somatic hypermutation takes place at
this time, no high affinity cells may emerge.
When the mutation begins after $3$ days, the number of
high affinity B-cells grows in a controlled way. It then
turns to a steep increase and finally reaches a plateau
on a high level for the last days of the GC reaction.
In the following the relation of this general behavior to
some characteristic features of the GC reaction is
discussed.

\subsubsection{The role of recycling}

The importance of recycling for affinity maturation
has already been emphasized 
(Kepler \& Perelson, 1993; Oprea \et 2000).
Furthermore, it has been
shown using a dynamical model without space resolution
that a large recycling probability is necessary in order
to get a reliable optimization process 
(Meyer-Hermann \et 2001).
As has already been mentioned,
the model considers those B-cells as recycled which have been
positively selected, at least once, in interaction with FDCs and which
restart to proliferate. This does not
mean that these cells reenter the dark zone. 
On the contrary, the model
predicts that recycling occurs in the light
zone and -- most interestingly -- that recycled cells remain in the light
zone and do not return to the dark zone, as was suspected
several times
(Kelsoe, 1996; Han \et 1997).
This can clearly be seen in Fig.~\ref{cbcc0x29a}:
the recycled cells all remain in the light zone.

\begin{figure}[ht]
\begin{center}
\includegraphics[height=11cm]{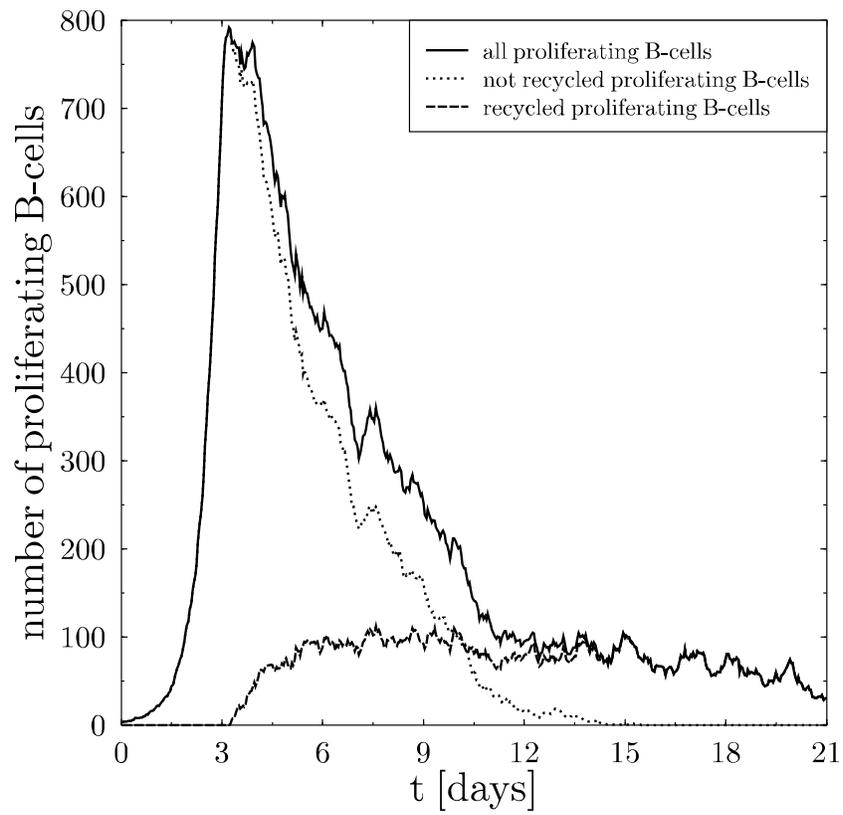}
\end{center}
\vspace*{-5mm}
\caption[]{\sf The time course of the whole population of 
centroblasts, the subset of recycled centroblasts, 
and non-recycled centroblasts
in the GC reaction.
The parameters in Tab.~\ref{parameter} were used.}
\label{0x29acb}
\end{figure}

The kinetics of the fraction of non-recycled and recycled B-cells 
is shown in Fig.~\ref{0x29acb}. 
In the beginning the population of
proliferating B-cells is dominated by 
non-recycled B-cells, and
recycling does not play an important role.
After $10.0\pm0.5$ 
days the proliferating cells are composed by equal
parts of non-recycled and recycled B-cells.
After $13.7\pm1.2$ 
days the non-recycled B-cell population vanishes.
Afterwards, all proliferating B-cells are recycled cells,
i.e.~they are stemming from at least one positive selection.

The interpretation of this result is quite obvious. Until day $10$ of 
the GC reaction an important part of the optimization procedure
has already taken place (see Fig.~\ref{0x29aaff}). 
This is in accordance with the
experimental observation that at this time the GC is already
dominated by good cells (Jacob \et 1993).
As a consequence, recycling does not seem to be of great
importance for the affinity maturation process
at this time. If one stopped the GC reaction at this
time, the cell population would have a higher average affinity to the
antigen than the seeder cells. Nevertheless, the number
of {\it optimal} cells would still be small and also a lot of low
affinity cells would remain in the GC.
The situation completely changes afterwards, when
the dark zone has disappeared, and non-recycled B-cells
become dominated by recycled B-cells. 
At this stage no low affinity cells 
are present in the GC anymore and there remains a slowly
decreasing population of already positively selected B-cells.

\begin{figure}[ht]
\begin{center}
\includegraphics[height=11cm]{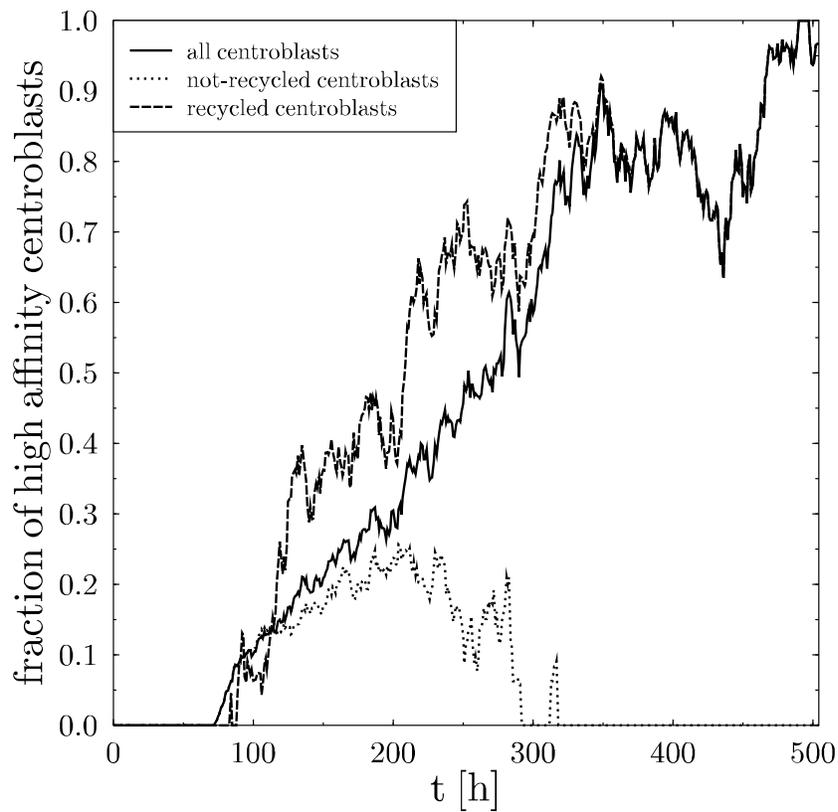}
\end{center}
\vspace*{-5mm}
\caption[]{\sf The time course of the fraction of high affinity
centroblasts in the GC reaction
(cells which bind to FDCs with a probability of at least $30\%$)
shown for the total centroblast population and
for the subsets of recycled B-cells and non-recycled centroblasts
separately. The fractions are defined with respect to the
total population of the corresponding subset (e.g. high affinity
recycled B-cells divided by all recycled B-cells).
The parameters in Tab.~\ref{parameter} were used.}
\label{0x29acbh}
\end{figure}

Note that the average number of mutations of B-cells 
that differentiate into output cells is about $3$ when the
dark zone vanishes and about $9$ at the end of the GC reaction
-- which is in accordance with experiment 
(K\"uppers \et 1993; Wedemayer \et 1997).
It follows that about $2/3$ of
all mutations in B-cells, that do not die by apoptosis but
are selected and differentiate into output cells
later on, occur in recycled B-cells. 
As can be seen in Fig.~\ref{0x29acbh},
these late mutations are not useless fluctuations of the affinity
but still substantially increase the affinity to the antigen.
This reveals that a major
part of the affinity maturation process is realized during
domination of the recycled B-cells.
Consequently, recycling is not only the basis
for the multiplication of those cells which are
already of relatively high quality but also for the fine tuning of these
cells which appears to be a relevant part of 
the affinity maturation process.

\subsubsection{The role of the dark zone}

\begin{figure}[ht]
\begin{center}
\includegraphics[height=8cm]{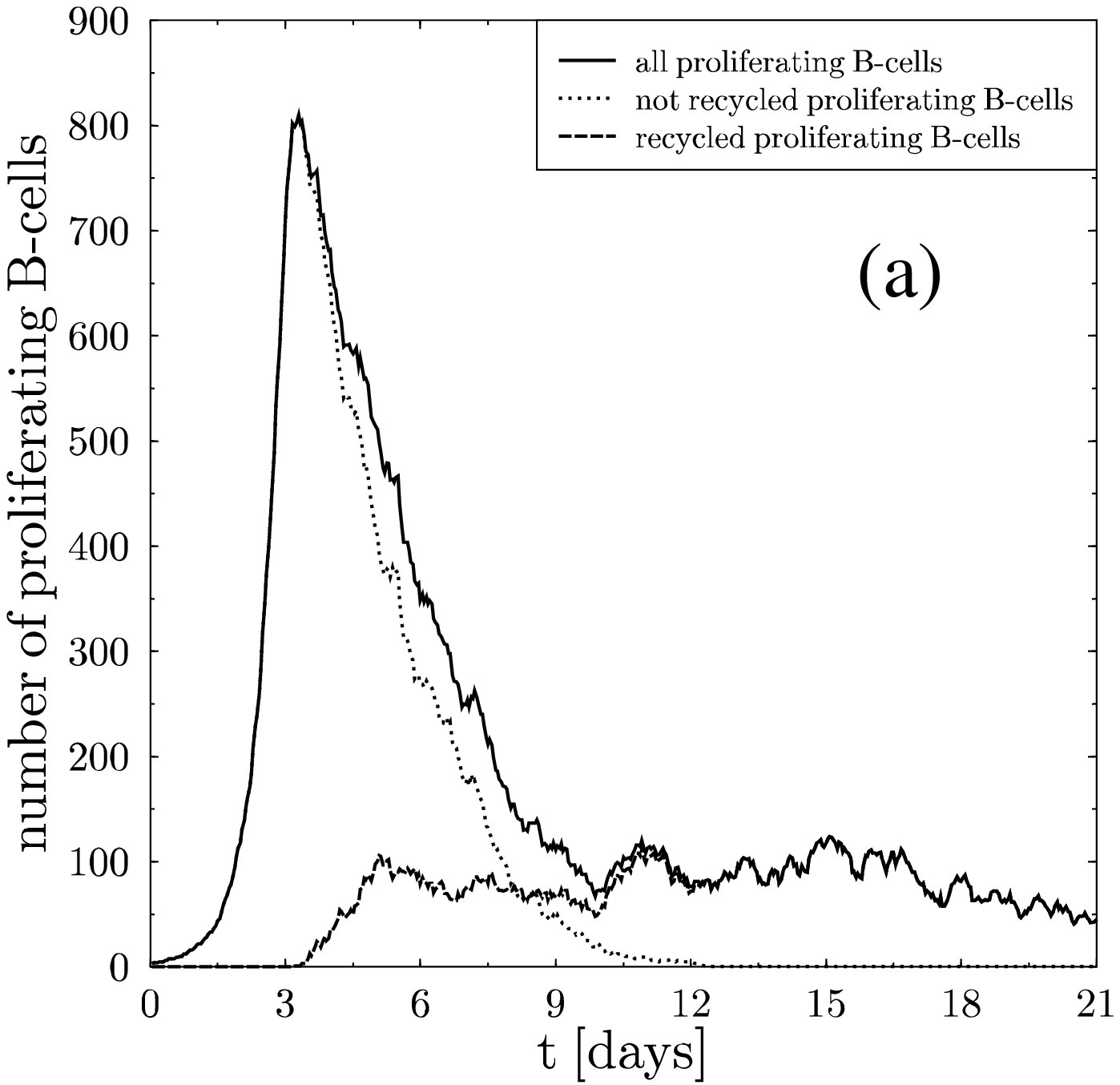}
\includegraphics[height=8cm]{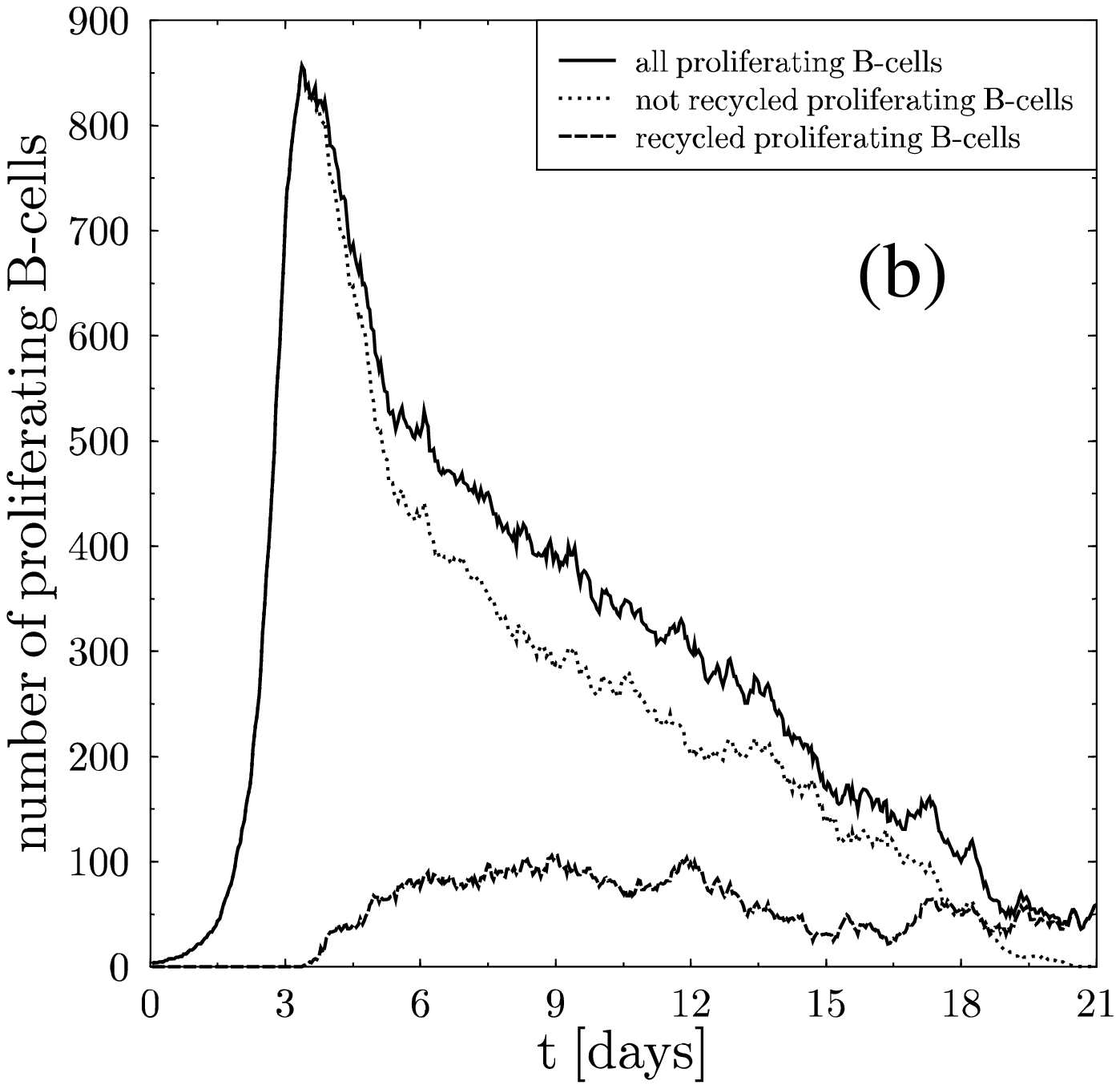}\\
\includegraphics[height=8cm]{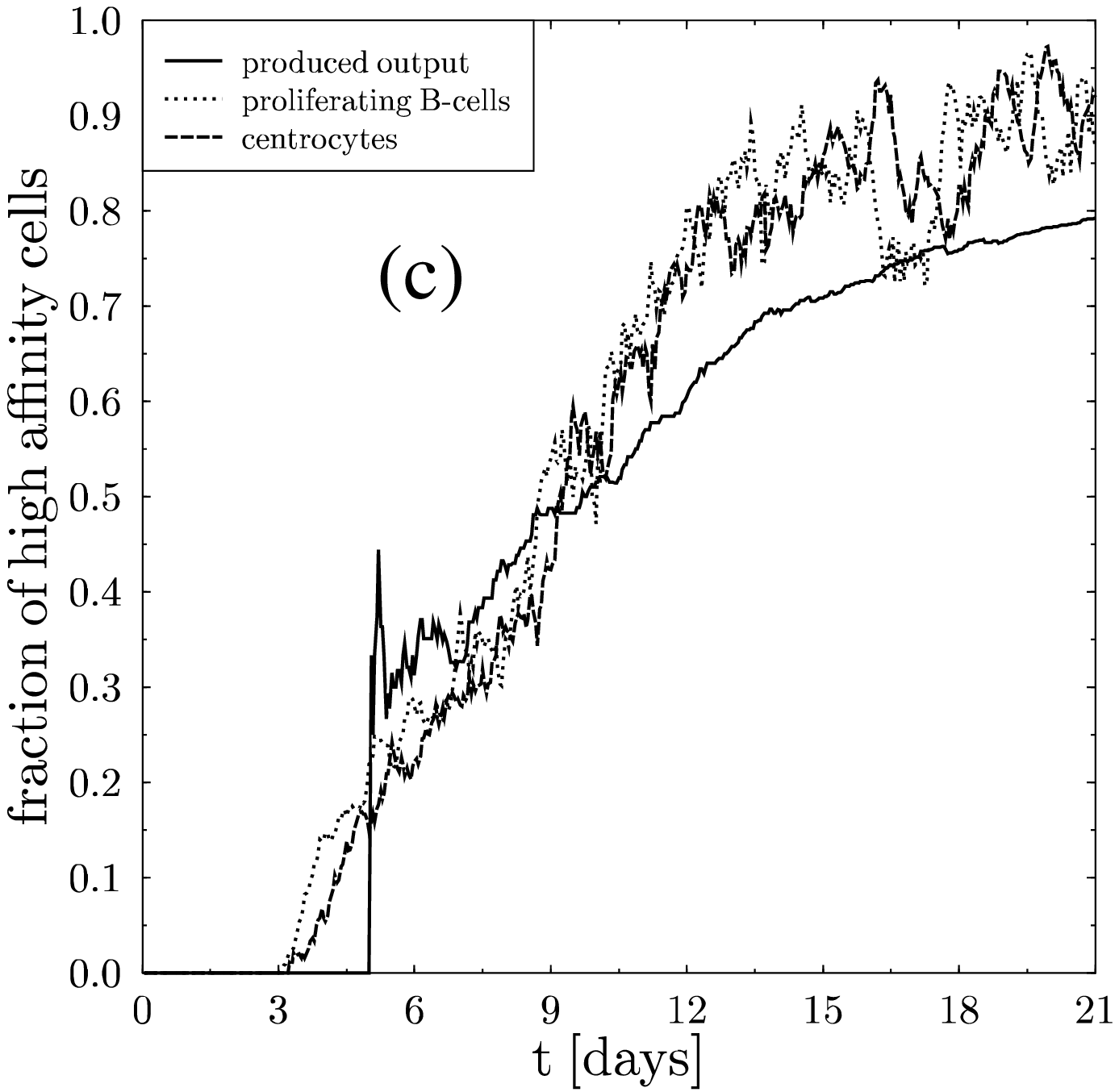}
\includegraphics[height=8cm]{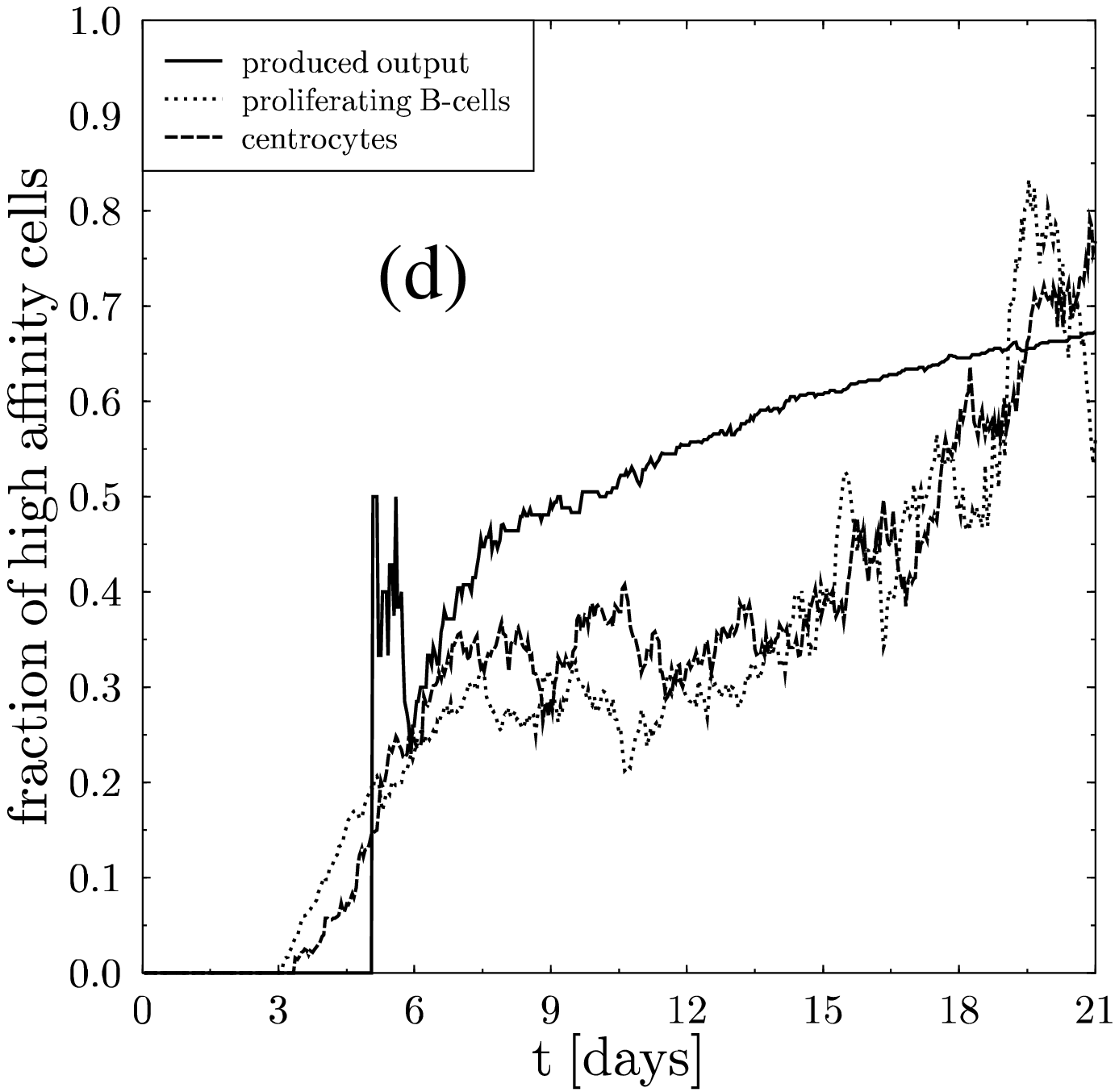}
\end{center}
\vspace*{-5mm}
\caption[]{\sf The time courses of the populations of 
non-recycled centroblasts, recycled centroblasts, and the sum
of all proliferating B-cells,
as well as of the fractions of high affinity B-cells. 
The duration of the dark zone was changed with respect
to Fig.~\ref{0x29aaff} and \ref{0x29acb} by the use
of different differentiation signal productions rates of 
$s=12/hr$ in (a,c) and $s=6/hr$ in (b,d). The apoptosis rate
is $z=1/(8\,hr)$ in (b,d).
All other parameters are like in Tab.~\ref{parameter}.
One observes a clear correlation between the moment
of domination of recycled B-cells (see (a) and (b)) and
the starting time of a phase of steep increase in the affinity
(see (c) and (d)).}
\label{cbandaff}
\end{figure}

One can observe an interesting correlation between the moment
when recycled B-cells dominate over non-recycled centroblasts
in the GC and the onset of the steep growth phase 
(mentioned at the beginning of this section) of the average cell
affinity to the antigen (compare Fig.~\ref{0x29aaff} and
\ref{0x29acb}, \ref{cbandaff} (a) and (c), 
as well as \ref{cbandaff} (b) and (d)).
The domination of recycled B-cells has to be preceded
by the destruction of the dark zone which is indeed
the case. Therefore, the phase of accelerated growth of average
affinity in the GC is not primarily due to an accelerated
affinity maturation but to the elimination of
low affinity cells which still proliferate in the
dark zone. Note that when the non-recycled centroblasts are
completely eliminated from the GC, the average cell affinity
already reaches a plateau.

As a second interesting observation, the
affinity development of all produced output cells
and of centroblasts and centrocytes differs considerably.
Before the steep increase in average affinity the
output cells are of better quality than the proliferating
B-cells -- afterwards the proliferating B-cells are
better or at least equally good (see Fig.~\ref{cbandaff}
(c), and (d)). Note that the
affinity of centroblasts and centrocytes has to become
larger than the affinity of the already produced 
output cells. Otherwise, a further increasing quality 
of all produced output cells together becomes impossible. 

Taking both observations together one is led to the
following hypothesis:
The affinity gap between the cells in the dark
and the light zone becomes larger during the GC reaction,
and in the late phase of the GC reaction
a further enhancement of the total affinity is inhibited by the
low affinity cells in the dark zone. 
Consequently, the dark zone has to be dissolved in time
in order to optimize the affinity maturation process.

If this hypothesis holds true, a correlation should exist
between the fraction of high affinity output cells 
(averaged over the whole GC reaction) 
and the duration of the dark zone.
This can be studied by varying the production
rate $s$ for the centroblast differentiation signal molecule, 
which is one of the parameters that control the life span
of the dark zone.
The parameters in Tab.~\ref{parameter} are used and
the signal production rate is varied in the range of
$6/hr\le s \le \infty$, where $\infty$ implies that the centroblasts
receive the signal without any time delay. 
Since the final number of B-cells 
is influenced from these changes, the centroblast
differentiation rate is adapted correspondingly:
$5.5/hr\le g_1 \le 6.4/hr$.
This ensures to find (in average) the same number of cells at the
end of the GC reaction.

\begin{figure}[ht]
\begin{center}
\includegraphics[height=8cm]{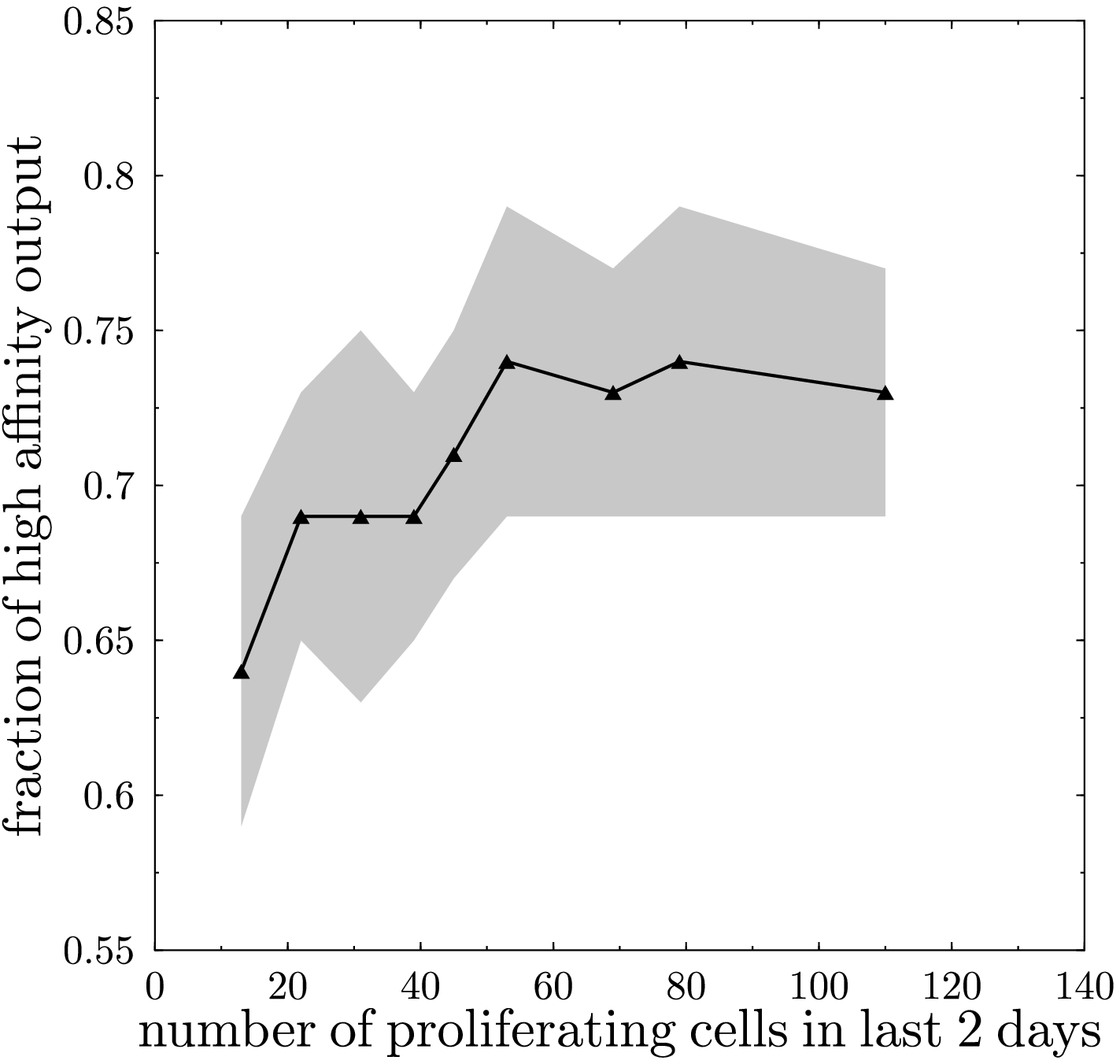}
\includegraphics[height=8cm]{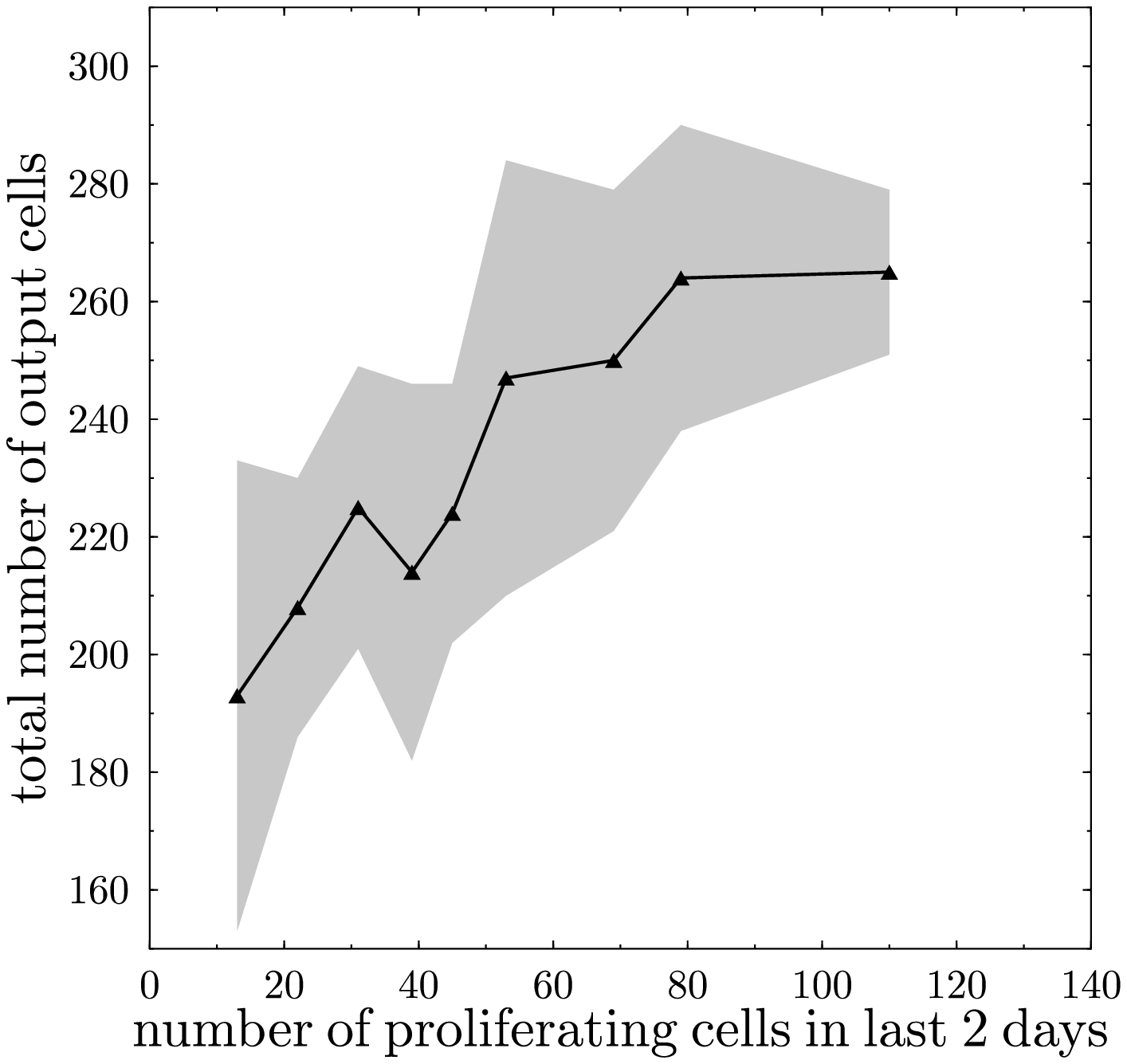}
\end{center}
\vspace*{-5mm}
\caption[]{\sf The dependence of the fraction of high affinity
output cells and of the total number of produced output cells
(summed up to the end of the GC reaction)
on the number of proliferating cells that remain
in average in the GC during the last $2$ days of the reaction. 
The grey area denotes one standard deviation of the average 
values along the full line.
The parameters in Tab.~\ref{parameter} were used
with different rates of centroblast differentiation signal production $s$,
i.e.~with dark zones of variable duration. The centroblast differentiation
rate $g_1$ is adapted correspondingly.}
\label{stacbqua}
\end{figure}

\begin{figure}[ht]
\begin{center}
\includegraphics[height=8cm]{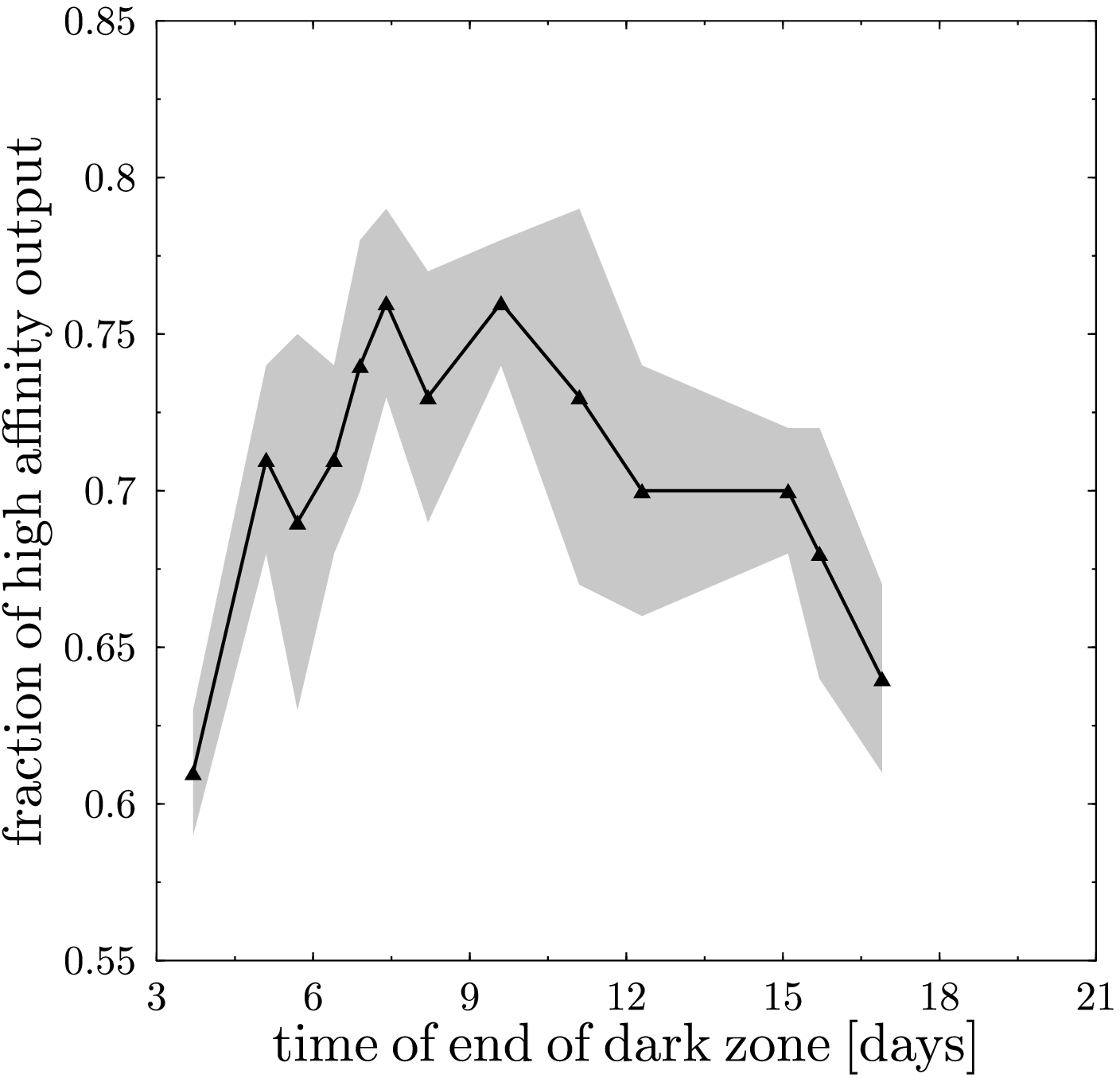}
\includegraphics[height=8cm]{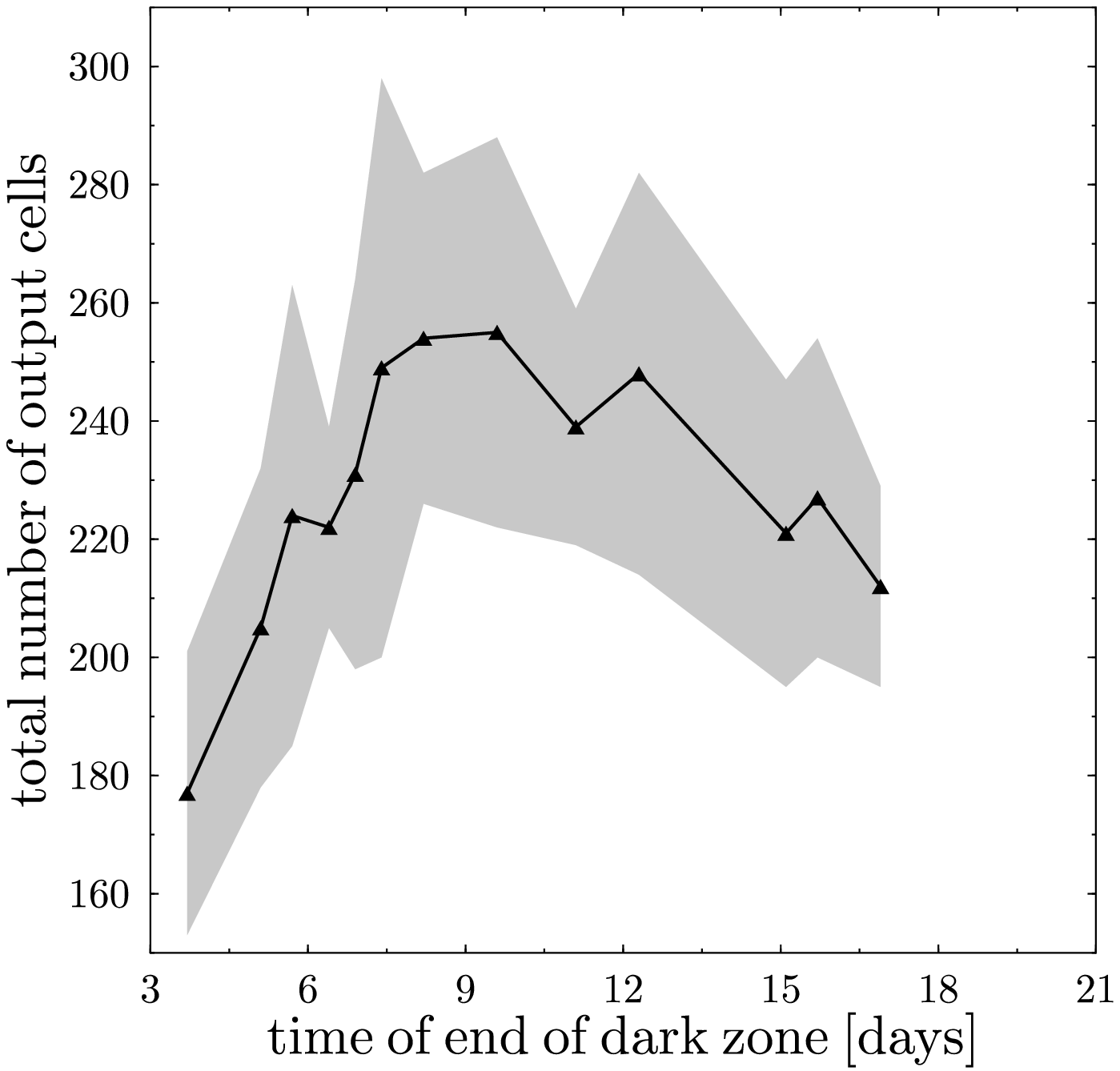}
\end{center}
\vspace*{-5mm}
\caption[]{\sf The dependence of the fraction of high affinity
output cells and of the total number of produced output cells 
(summed up to the end of the GC reaction)
on the duration of the dark zone. 
The grey area denotes one standard deviation of the average 
values along the full line.
The parameters in Tab.~\ref{parameter} were used, where
the production rate of the centroblast differentiation signal $s$ and
the centroblast differentiation rate $g_1$ were
varied in order to generate different durations of the dark zone.}
\label{staz}
\end{figure}

It is important to take care of comparability
of the simulations, as in a stochastic model
the quantitative results are variable
to some extent. This is especially true for the
final state of the GC reaction, i.e.~the number of
proliferating B-cells which remain in the GC at the end
of the reaction. 
As can be seen in Fig.~\ref{stacbqua},
the quality of the produced output cells (averaged over the
whole GC reaction) clearly correlates with the
final number of proliferating cells at day $21$ of the reaction
up to a certain level of those cells. Above that level the output
cell quality seems to saturate.
Therefore, an analysis concerning the
correlation between the length of the dark zone and
the output-quality has to be based on simulations with
comparable final states, in order not to hide one
correlation with another one. Consequently, only those simulations
were taken into account that resulted in 
a reasonable number of remaining proliferating B-cells in the range of 
$50\pm 1\sigma$, where the standard deviation $\sigma$ is taken over
all simulations. The resulting dependency of the total output quality
(i.e.~averaged over the whole GC reaction)
on the duration of the dark zone is depicted in Fig.~\ref{staz}.
Indeed, the fraction of high affinity output cells decreases if
the dark zone remains stable for a too long period of more than
$10$ days. In conclusion, intermediate dark zones which vanish
before day $10$ of the GC reaction are 
advantageous for the affinity maturation process in GC
reactions.

On the other hand, Fig.~\ref{staz} shows that for too
short living dark zones that vanish before day $7$ of the reaction,
the quality of all produced output cells is again reduced. This leads
to the clear conclusion that the intermediate appearance
of dark zones optimizes the affinity maturation process
in GC reactions. The best results are found for dark zones
that vanish between day $7$ and $10$ of the GC reaction.

Finally, the number of output cells resulting
from the GC reaction is correlated with
the dark zone duration. The maximum number of output cells is
found for dark zones of exactly the same duration as that one
required from the point of view of cell quality
(see Fig.~\ref{staz}). 
Note, that the error bars are larger here than in the result
for the output quality.


\subsection{Evolution of the total volume}
\label{volume}

\begin{figure}[ht]
\begin{center}
\includegraphics[height=11cm]{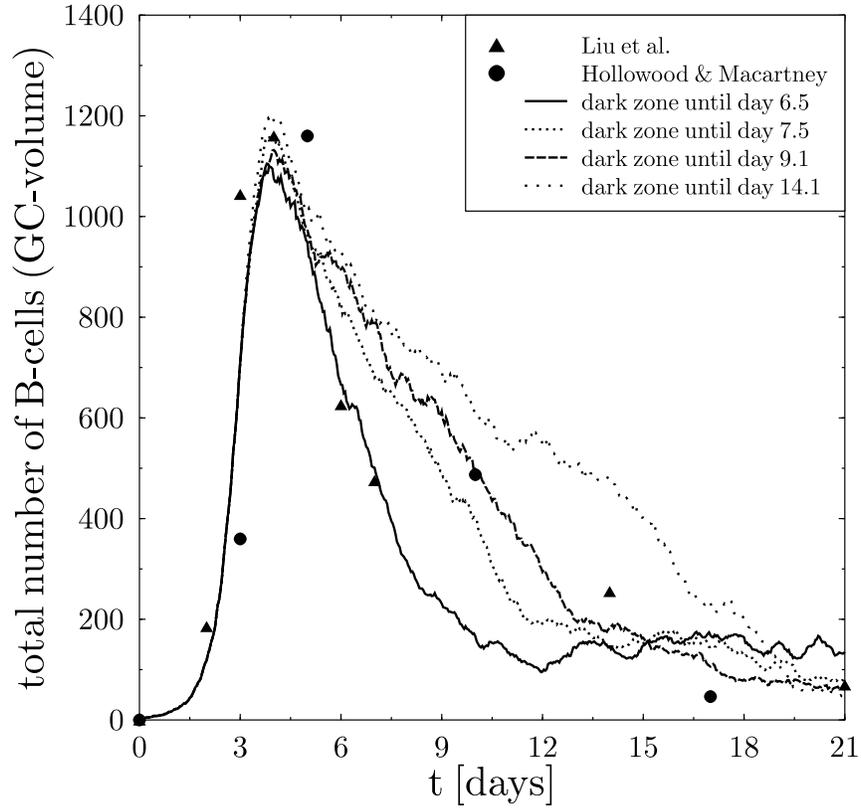}
\end{center}
\vspace*{-5mm}
\caption[]{\sf Representative time courses of the total number of B-cells 
(i.e.~centroblasts and centrocytes)
in the GC reaction for four different durations of the dark zone. 
The parameters in Tab.~\ref{parameter} were used for the
GC reactions with dark zones until day $7.5$ and $9.1$. 
The short dark zone was generated with a faster differentiation
signal production of $s=12/hr$.
In the case of the long dark zone
the centroblast differentiation signal is produced with a
lower rate of $s=6/hr$ and the centroblast differentiation rate
is $g_1=1/(5.65\,hr)$. For comparison, the average data from
(Liu \et 1991; Hollowood \& Macartney, 1992) 
were normalized to the maximum of the
simulation results and rescaled correspondingly.}
\label{GCvolume}
\end{figure}

The evolution of the total GC volume during the GC reaction
after primary immunization has been observed in experiment 
by measuring the follicle center volume in per cent of the total 
splenic volume (Liu \et 1991) or the total GC volume
itself (Hollowood \& Macartney, 1992).
In both cases a statistical average over a lot
of GCs found in the spleen was performed.
These measurements are suitable for comparison with the present model,
even if the absolute numbers cannot be interpreted.
The observed time course 
has the following properties: 
At first, the splenic GC volume increases steeply. It reaches
a maximum at day $4$ (Liu \et 1991) or day $5$ (Hollowood \& Macartney, 1992)
of the reaction, and, subsequently, decreases until the end 
of the reaction at day $21$.
This qualitative behavior is perfectly reproduced by the model
for intermediate dark zones that vanish between day $6.5$ and $9.1$ of the GC
reaction (see Fig.~\ref{GCvolume}). 
As in experiment,
the model shows an exponential increase of the GC volume,
that is slowing down around day $3$ of the reaction and reaches its
maximum at day $4$. Finally, a fast decrease
of the volume follows becoming more moderate in the late phase of the
reaction.
An interesting observation is that
the time course of GC reactions with longer dark zones 
until day $14.1$ quantitatively is not in accordance with the data
found in experiment (see Fig.~\ref{GCvolume}).
This is another indication for an intermediate and relative
short appearance of dark zones in GC reactions.

However, one important question remains unresolved in this
context: How does the GC reaction end? In the framework
of this model no mechanism was assumed in order to stop
the reaction. The resulting final state of the reaction is a
strongly decreased but, nevertheless, relatively stable
population of B-cells. Therefore, we are led to the question
if, eventually, there exists an additional mechanism that
terminates the GC reaction.

Comparing the GC-volume with the already
mentioned experiment (Liu \et 1991) we can find a first
hint. It seems that in the last phase of the GC reaction
the volume remains on a higher level than in the model
and decreases faster at the very end (see Fig.~\ref{GCvolume}).
This behavior may be achieved in the model
e.g.~with a slower selection procedure which stabilizes the total
B-cell population on a higher level.
However, the model does not provide any possibility for 
the rapid population decrease during the last days, so that
the reaction would remain on a high level after $21$ days.
This would have to be introduced separately, e.g.
by incorporating a change of the proliferation rate or the
recycling probability during the GC reaction.
The latter one may be favored: 
It is believed that the differentiation
process of positively selected centrocytes is determined
by a local interaction with FDCs. It would not
be very surprising if the proportion of the signals
that induce recycling and differentiation into output cells
changed during the GC reaction, for example in dependence
on the already achieved level of affinity to the antigen.

In this context it is interesting to note that the total number
of produced output cells (summed over the whole GC reaction)
becomes larger if more proliferating B-cells remain in the GC 
during its last phase (see Fig.~\ref{stacbqua}).
This correlation indicates that a prolonged high number
of B-cells that decreases rapidly at the very end of the 
reaction may be advantageous for the
quantity of the GC output.

\subsection{Robustness of the results}
\label{robust}

\subsubsection{Stochastic variation}
\label{stochastic}

One should be aware of the fact that all simulations were based
on the same FDC and seeder cell 
configuration. This has been necessary
in order to uncover the GC features and dependencies on the
model parameters. If the simulation is started by a stochastic
process from the very beginning of the GC reaction the variability
of the results becomes substantially larger. Especially, the
specific correlations during the dynamic development of the
GC reaction would partly be hidden by the variability of the
state after $3$ days of pure monoclonal expansion of
centroblasts. This state is very sensitive to the initial cell
distribution and to changes of 
the generator of random numbers, as the number of cells is
very low during the first hours of the reaction. Therefore,
slight differences at the beginning are crucial for the
development and blow up when the GC further expands.

This restriction of stochasticity was useful, as the main
interest was an analysis of the main part of the GC reaction.
However, the variability of real healthy GCs may be substantially
larger than shown here.

\subsubsection{Affinity of seeder cells}

As mentioned in Sec.~\ref{phases} the seeder B-cells that initiate the GC reaction
have a low but non-vanishing affinity to the antigen. In the presented
simulation runs they have to
mutate at least $5$ times in order to find the optimal mutant. In this
case one gets a rather stable answer measured in terms of the quality of the
resulting cells. However, the number of necessary mutations starting from the 
initial state is crucial for the GC development. If the affinity of the seeder
cells is too low, the optimal mutant remains unreached in the framework
of the model. Consequently, the initial distribution of the seeder cells
in the shape space is important for the outcome of the GC reaction
(see also Meyer-Hermann et al., 2001).

It has been observed in experiments that GCs developing during an immune
response can be divided into two classes: Either high affinity cells remain
practically absent or they already dominate in an early stage of the reaction
(Berek et al., 1991; Radmacher et al., 1998).
This behavior may correspond to the above mentioned limits of high
and low affinity seeder cells. It is the aim of current statistical investigation to
decide if the model reproduces a sharp switch between the two 
observed GC classes or not.

\subsubsection{Selection process}
\label{robust_select}

The affinity maturation process is basically governed
by the centrocyte life time, the time
a centrocyte needs to bind to an FDC, and the {\it total}
selection time. 
The model results are indeed insensitive to a complementary
change of the duration of different selection substate, i.e.~one
may increase the centrocyte differentiation rate and reduce the rate
of centrocyte-FDC-dissociation, correspondingly. But the results
are sensitive to a change of only one of these parameters
or of the apoptosis rate.
We conclude that in the framework of the model 
the duration of the different selection substates
(starting from the binding of centrocytes and FDCs and 
ending with the differentiated centrocyte)
is not of direct relevance for the affinity maturation process.
However, the total selection time turned out to be important.


\subsubsection{Diffusion constants}
\label{robdiffuse}

The diffusion constants -- besides the availability
of space -- determine the mobility of the B-cells in the GC volume.
Unfortunately, their values for centroblasts and centrocytes
are rather uncertain. Therefore, it is important to check
the relevance of these parameters for the characterization
of the GC reaction. If the mobility of the B-cells is
larger, the total number of B-cells is increased in the
late phase of the GC reaction. The B-cells get a higher
chance of reaching an FDC and, in this way, have a higher
chance of being selected. This becomes especially relevant
in the late phase of the GC reaction as the density of
cells becomes small and their mobility plays a major role.
It is not surprising that for smaller diffusion constants
the total number of cells is reduced in the late phase.
Since the total number of produced output cells is directly related
to the final number of B-cells (see Fig.~\ref{stacbqua})
the produced output should
not be interpreted in terms of absolute numbers.

The affinity maturation process is slightly influenced by
the diffusion constants. A faster diffusion basically has the
same effect as a slower apoptosis rate.
However, the relation of the affinity maturation process 
to the dark zone is not altered.

The outcome of the simulations does not depend on the
diffusion constant of the signal molecule until the values
remain substantially larger than the diffusion constant
of the B-cells.


\section{Discussion}
\label{discuss}

The aim of this article was to reveal the origin and the function
of the specific morphology of the GC reaction. Concerning the
origin, two requirements were found: 
It is essential for the development of dark zones that 
during the first phase of monoclonal expansion
the centroblasts also occupy a volume besides the FDC network.
Secondly, it was shown that a dark zone develops if 
the centroblast differentiation to centrocytes is governed by a
signal molecule, which is produced by 
the FDCs and consumed by the centroblasts. This is surely not
the only possible scenario. However, the non-local character of the
interaction between cells outside the dark zone and centroblasts 
is an essential ingredient. 
It has been shown that other obvious concepts for
the centroblast differentiation lead to contradictions with
experimental observations. 
For example, 
centroblast-specific differentiation does not allow for the development
of dark zones at all, even with a centrocyte attractor in the FDC network.
A local centroblast-FDC interaction leads to a dominance of centroblasts
that is completely untypical in real GC reaction. This
concept may be restored if one introduces an early decreasing
proliferation rate, which again is not in agreement with experimental
data. 

The concept of non-local centroblast-FDC interaction is compatible
with dark zones of different durations, i.e.~with stable or
intermediate dark zones. Intermediately appearing
dark zones have been observed in experiment (Camacho \et 1998). In order to
understand the function of the dark zone, the quality of the
B-cells resulting from the GC reaction with respect
to the antigen, i.e.~the quality of the affinity maturation process,
has been examined. We showed that
the affinity maturation process is inhibited
if the dark zone does not vanish until day $10$ of 
the GC reaction.
Shorter dark zones lead to a better
optimization process and, consequently, to plasma- and memory-cells
of higher quality.
On the other hand, the dark zone has to exist until day $7$ of
the GC reaction in order to optimize the affinity maturation
process. 
Also, the quantity of produced plasma- and memory-cells
were found to be maximized for dark zones of exactly this duration.
These results suggest that the dark zone has the function to
present a large pool of different B-cells with a great diversity
to the FDCs for selection, and at the same time to increase the total number
of resulting antibody-producing cells. Therefore, the
development of a dark zone as well as its depletion in time
are essential elements of the affinity maturation process.

The time course of the GC volume predicted by the model was
compared to experimentally observed GC volumes.
The model results are in agreement with experiment for
dark zones that vanish between day $6.5$ and $9.1$ of the
GC reaction. This supports the above argument for the
existence of an intermediate dark zone, and suggests
that nature has chosen to realize GCs with exactly this
property. So the main conclusion of this analysis is
that affinity maturation is optimized for intermediate 
dark zones that vanish around day $8$ of the GC reaction, 
and that it seems
that this optimal procedure has been realized in existing
GCs.

In accordance with previous works,
recycling, i.e.~the process that centrocytes re-proliferate,
turned out to be essential for a well working affinity maturation.
More explicitly, the
affinity maturation process is divided into a phase of
rough optimization which is dominated by B-cells in the
dark zone and into a phase of fine tuning which is dominated by
recycled B-cells. It is worth emphasizing in this context, 
that the number of recycled B-cells that reenter the dark
zone is negligible.

At this point one should think about the optimization
of the model. The present model is in accordance with
experimentally observed GCs, especially including
quantitative statements. Also the time course of the
GC volume has been found to be in agreement with experiment
and supports the hypothesis of an
intermediate dark zone of relatively short duration.
The total GC volume is reduced to a reasonable small
number of remaining cells without any changes in the
dynamical parameters. 
However, 
the quality of the resulting plasma- and memory-cells is enhanced 
and their quality is increased
if the total number of B-cells remains larger than 
in the model during the last days of the GC reaction.
This points towards an additional mechanism
that stops the GC reaction at the very end of the reaction. 
It was argued, that a recycling probability which is
reduced depending on the already achieved affinity
maturation is a good candidate. Also the production of
antibodies inside the GC and the implied binding of antigen
fragments may be considered in this context.

Finally, the role of T-helper cells should be investigated.
They are believed to have an important influence on
the selection process as well as on the inhibition
of apoptosis (Lindhout \et 1995; Brandtzaeg, 1996; 
Tew \et 1997; Hollmann \& Gerdes, 1999;
Hur \et 2000; van Eijk \et 2001). 
Therefore, T-helper cells should
be considered if one is interested in the details
of the selection process, which did not turn out to be
of primary importance for the development of the dark zone
analysed in the present article.
In addition, T-helper cells instead of the FDCs 
are also a good candidate to produce
the centroblast differentiation signal 
(Dubois \et 1999a, 1999b), that has been postulated here.



\vfill
\eject
\newcounter{fig}
\section*{References}
\begin{list}{\normalsize
{\rm }}
{\usecounter{fig}
\setcounter{fig}{0}
\labelwidth0mm
\leftmargin8mm
\rightmargin0mm
\labelsep0mm
\topsep0mm
\parsep0mm
\itemsep0mm}
\itemindent-8mm
\normalsize

\item
{\sc Berek, C.~\& Milstein, C.},
{\rm 1987}:
{\rm Mutation drift and repertoire shift in the maturation 
of the immune response},
{\it Immunol. Rev.\/}
{\bf 96},
{\rm 23-41}.

\item
{\sc Berek, C., Berger, A.~\& Apel, M.},
{\rm 1991}:
{\rm Maturation of the Immune Response In Germinal Centers},
{\it Cell\/}
{\bf 67},
{\rm 1121-1129}.

\item
{\sc Berek, C.},
{\rm 2001}:
{\rm private communication}.

\item
{\sc Beyer, T.~\& Meyer-Hermann, M.},
{\rm 2002}:
{\rm A corresponding work is currently in preparation}.

\item
{\sc Brandtzaeg, P.},
{\rm 1996}:
{\rm The B-cell development in tonsillar lymphoid follicles},
{\it Acta Otolaryngol. Suppl. (Stockh)\/}
{\bf 523},
{\rm 55-59}.

\item
{\sc Camacho, S.A., Koscovilbois, M.H.~\& Berek, C.},
{\rm 1998}:
{\rm The Dynamic Structure of the Germinal Center},
{\it Immunol. Today\/}
{\bf 19},
{\rm 511-514}.

\item
{\sc Choe, J.~\& Choi, Y.S.},
{\rm 1998}:
{\rm IL-10 Interrupts Memory B-Cell Expansion in the Germinal 
Center by Inducing Differentiation into Plasma-Cells},
{\it Eur. J. Immunol.\/}
{\bf 28},
{\rm 508-515}.

\item
{\sc Choe, J., Li, L., Zhang, X., Gregory, C.D.~\& Choi, Y.S.},
{\rm 2000}:
{\rm Distinct Role of Follicular Dendritic Cells and T Cells 
in the Proliferation, Differentiation, and Apoptosis 
of a Centroblast Cell Line, L3055},
{\it J. Immunol.\/}
{\bf 164},
{\rm 56-63}.

\item
{\sc Dubois, B., Barth\'el\'emy, C., Durand, I., Liu, Y.-J., Caux, 
C.~\& Bri\`ere, F.},
{\rm 1999a}:
{\rm Toward a Role of Dendritic Cells in the Germinal Center 
Reaction -- Triggering of B-Cell Proliferation and Isotype 
Switching},
{\it J. Immunol.\/}
{\bf 162},
{\rm 3428-3436}.

\item
{\sc Dubois, B., Bridon, J.-M., Fayette, J., Barth\'el\'emy, 
C., Banchereau, J., Caux, C.~\& Bri\`ere, Francine},
{\rm 1999b}:
{\rm Dendritic Cells directly modulate B cell growth and 
differentiation},
{\it J. Leukoc. Biol.\/}
{\bf 66},
{\rm 224-230}.

\item
{\sc Dunn-Walters, D., Thiede, C., Alpen, B.~\& Spencer, J.},
{\rm 2001}:
{\rm Somatic hypermutation and B-cell lymphoma},
{\it Philos. Trans. R. Soc. Lond. B Biol. Sci.\/}
{\bf 356},
{\rm 73-82}.

\item
{\sc Falini, B. et al.},
{\rm 2000}:
{\rm A monoclonal antibody (MUM1p) detects expression of the MUM1/IRF4 
protein in a subset of germinal center B cells, plasma cells, and activated
T cells},
{\it Blood\/}
{\bf 95},
{\rm 2084-2092}.

\item
{\sc Han, S.H., Zheng, B., Dal Porto, J.~\& Kelsoe, G.},
{\rm 1995a}:
{\rm In situ Studies of the Primary Immune Response to (4-Hydroxy-3-Nitrophenyl) 
Acetyl IV. Affinity-Dependent, Antigen-Driven B-Cell 
Apoptosis in Germinal Centers as a Mechanism for Maintaining 
Self-Tolerance},
{\it J. Exp. Med.\/}
{\bf 182},
{\rm 1635-1644}.

\item
{\sc Han, S.H., Hathcock, K., Zheng, B., Kelper, T.B., Hodes, 
R.~\& Kelsoe, G.},
{\rm 1995b}:
{\rm Cellular Interaction in Germinal Centers: Roles of 
CD40-Ligand and B7-1 and B7-2 in Established Germinal 
Centers},
{\it J. Immunol.\/}
{\bf 155},
{\rm 556-567}.

\item
{\sc Han, S., Zheng, B., Takahashi, Y.~\& Kelsoe, G.},
{\rm 1997}:
{\rm Distinctive characteristics of germinal center B cells},
{\it Immunology\/}
{\bf 9},
{\rm 255-260}.

\item
{\sc Hanna, M.G.},
{\rm 1964}:
{\rm An autoradiographic study of the germinal center in 
spleen white pulp during early intervals of the immune 
response},
{\it Lab. Invest.\/}
{\bf 13},
{\rm 95-104}.

\item
{\sc Hollmann, C.~\& Gerdes, J.},
{\rm 1999}:
{\rm Follicular Dendritic Cells and T-Cells -- Nurses and 
Executioners in the Germinal Center Reaction},
{\it J. Pathol.\/}
{\bf 189},
{\rm 147-149}.

\item
{\sc Hollowood, K.~\& Macartney, J.},
{\rm 1992}:
{\rm Cell kinetics of the
germinal center reaction -- a stathmokinetic study},
{\it J. Immunol.\/}
{\bf 22},
{\rm 261-266}.

\item
{\sc Hostager, B.S., Catlett, I.M.~\& Bishop, G.A.},
{\rm 2000}:
{\rm Recruitment of CD40 and Tumor Necrosis Factor Receptor-associated 
Factors 2 and 3 to Membrane Microdomains during CD40 
Signaling},
{\it J. Biol. Chem.\/}
{\bf 275},
{\rm 15392-15398}.

\item
{\sc Hur, D.H., Kim, D.J., Kim, S., Kim, Y.I., Cho, D., Lee, 
D.S., Hwang, Y.-I., Bae, K.-W., Chang, K.Y.~\& Lee, W.J.},
{\rm 2000}:
{\rm Role of follicular dendritic cells in the apoptosis 
of germinal center B cells},
{\it Immunol. Lett.\/}
{\bf 72},
{\rm 107-111}.

\item
{\sc Jacob, J., Kassir, R.~\& Kelsoe, G.},
{\rm 1991}:
{\rm In situ studies of the primary immune response to 
(4-hydroxy-3-nitrophenyl)acetyl. 
I. The architecture and dynamics of responding cell 
populations},
{\it J. Exp. Med.\/}
{\bf 173},
{\rm 1165-1175}.

\item
{\sc Jacob, J., Przylepa, J., Miller, C.~\& Kelsoe, G.},
{\rm 1993}:
{\rm In situ studies of the primary response to (4-hydroxy-3-nitrophenyl)acetyl. 
III. The kinetics of V region mutation and selection 
in germinal center B cells},
{\it J. Exp. Med.\/}
{\bf 178},
{\rm 1293-1307}.

\item
{\sc Kelsoe, G.},
{\rm 1996}:
{\rm The germinal center: a crucible for lymphocyte selection},
{\it Semin. Immunol.\/}
{\bf 8},
{\rm 179-184}.

\item
{\sc Kepler, T.B.~\& Perelson, A.S.},
{\rm 1993}:
{\rm Cyclic re-entry of germinal center B cells and the 
efficiency of affinity maturation},
{\it Immunol. Today\/}
{\bf 14},
{\rm 412-415}.

\item
{\sc Kesmir, C.~\& de Boer, R.J.},
{\rm 1999}:
{\rm A Mathematical Model on Germinal Center Kinetics and 
Termination},
{\it J. Immunol.\/}
{\bf 163},
{\rm 2463-2469}.

\item
{\sc Koopman, G., Keehnen, R.M., Lindhout, E., Zhou, D.F., de 
Groot, C.~\& Pals, S.T.},
{\rm 1997}:
{\it Eur. J. Immunol.\/}
{\bf 27},
{\rm 1-7}.

\item
{\sc Kroese, F.G., Wubbena, A.S., Seijen, H.G.~\& Nieuwenhuis, P.},
{\rm 1987}:
{\rm Germinal centers develop oligoclonally},
{\it Eur. J. Immunol.\/}
{\bf 17},
{\rm 1069-1072}.

\item
{\sc K\"uppers, R., Zhao, M., Hansmann, M.L.~\& Rajewsky, K.},
{\rm 1993}:
{\rm Tracing B Cell Development in Human Germinal Centers 
by Molecular Analysis of Single Cells Picked from Histological 
Sections},
{\it EMBO J.\/}
{\bf 12},
{\rm 4955-4967}.

\item
{\sc Lindhout, E., Lakeman, A.~\& de Groot, C.},
{\rm 1995}:
{\rm Follicular dendritic cells inhibit apoptosis in human 
B lymphocytes by rapid and irreversible blockade of 
preexisting endonuclease},
{\it J. Exp. Med.\/}
{\bf 181},
{\rm 1985-1995}.

\item
{\sc Lindhout, E., Koopman, G., Pals, S.T.~\& de Groot, C.},
{\rm 1997}:
{\rm Triple check for antigen specificity of B cells during 
germinal centre reactions},
{\it Immunol. Today\/}
{\bf 18},
{\rm 573-576}.

\item
{\sc Liu, Y.J., Barthelemy, C., De Bouteiller, O.~\& Banchereau, J.},
{\rm 1994}:
{\rm The differences in survival and phenotype between centroblasts 
and centrocytes},
{\it Adv. Exp. Med. Biol.\/}
{\bf 355},
{\rm 213-218}.

\item
{\sc Liu, Y.J., Joshua, D.E., Williams, G.T., Smith, C.A., Gordon, 
J.~\& MacLennan, I.C.},
{\rm 1989}:
{\rm Mechanism of antigen-driven selection in germinal centres},
{\it Nature\/}
{\bf 342},
{\rm 929-931}.

\item
{\sc Liu, Y.J., Zhang, J., Lane, P.J., Chan, E.Y.~\& MacLennan, 
I.C.M.},
{\rm 1991}:
{\rm Sites of specific B cell activation in primary and 
secondary responses to T cell-dependent and T cell-independent 
antigens},
{\it Eur. J. Immunol.\/}
{\bf 21},
{\rm 2951-2962}.

\item
{\sc Loeffler, M.~\& Stein, H.},
{\rm 2001}:
{\rm private communication}.

\item
{\sc MacLennan, I.C.M.},
{\rm 1994}:
{\rm Germinal Centers},
{\it Annu. Rev. Immunol.\/}
{\bf 12},
{\rm 117-139}.

\item
{\sc McHeyzer-Williams, M.G., McLean, M.J., Labor, P.A.~\& Nossal, 
G.V.J.},
{\rm 1993}:
{\rm Antigen-driven B cell differentiation in vivo},
{\it J. Exp. Med.\/}
{\bf 178},
{\rm 295-307}.

\item
{\sc Meyer-Hermann, M., Deutsch, A.~\& Or-Guil, M.},
{\rm 2001}:
{\rm Recycling Probability and Dynamical Properties of Germinal 
Center Reactions},
{\it J. Theor. Biol.\/}
{\bf 210},
{\rm 265-285}.

\item
{\sc Meyer-Hermann, M.},
{\rm 2002}:
{\rm Does Recycling in Germinal Centers exist?},
{\it Immunol. Cell Biol.}
{\bf 80},
{\rm 30-35}.

\item
{\sc Murray, J.D.},
{\rm 1993}:
{\rm Mathematical Biology}.
{\rm Springer},
{\rm Berlin u.a.},
{\rm 2.\,Aufl.}

\item
{\sc Nossal, G.},
{\rm 1991}:
{\rm The molecular and cellular basis of affinity maturation 
in the antibody response},
{\it Cell\/}
{\bf 68},
{\rm 1-2}.

\item
{\sc Oprea, M.~\& Perelson, A.S.},
{\rm 1996}:
{\rm Exploring the Mechanism of Primary Antibody Responses 
to T-Cell-Dependent Antigen},
{\it J. Theor. Biol.\/}
{\bf 181},
{\rm 215-236}.

\item
{\sc Oprea, M.~\& Perelson, A.S.},
{\rm 1997}:
{\rm Somatic mutation leads to efficient affinity maturation 
when centrocytes recycle back to centroblasts},
{\it J. Immunol.\/}
{\bf 158},
{\rm 5155-5162}.

\item
{\sc Oprea, M., van Nimwegen, E.~\& Perelson, A.S.},
{\rm 2000}:
{\rm Dynamics of One-pass Germinal Center Models: Implications 
for Affinity Maturation},
{\it Bull. Math. Biol.\/}
{\bf 62},
{\rm 121-153}.

\item
{\sc Pascual, V., Liu, Y.-J., Magalski, A., de Bouteiller, 
O., Banchereau, J.~\& Capra, J.D.},
{\rm 1994a}:
{\rm Analysis of somatic mutation in five B cell subsets 
of human tonsil},
{\it J. Exp. Med.\/}
{\bf 180},
{\rm 329-339}.

\item
{\sc Pascual, V., Cha, S., Gershwin, M.E., Capra, J.D.~\& Leung, 
P.S.C.},
{\rm 1994b}:
{\rm Nucleotide Sequence Analysis of Natural and Combinatorial 
Anti-PDC-E2 Antibodies in Patients with Primary Biliary 
Cirrhosis},
{\it J. Immunol.\/}
{\bf 152},
{\rm 2577-2585}.

\item
{\sc Perelson, A.S.~\& Oster, G.F.},
{\rm 1979}:
{\rm Theoretical Studies of Clonal Selection: Minimal Antibody 
Repertoire Size and Reliability of Self-Non-self Discrimination},
{\it J. Theor. Biol.\/}
{\bf 81},
{\rm 645-670}.

\item
{\sc Radmacher, M.D., Kelsoe, G.~\& Kepler, T.B.},
{\rm 1998}:
{\rm Predicted and Inferred Waiting-Times for Key Mutations 
in the Germinal Center Reaction -- Evidence for Stochasticity 
in Selection},
{\it Immunol. Cell Biol.\/}
{\bf 76},
{\rm 373-381}.

\item
{\sc Rundell, A., Decarlo, R., Hogenesch, H.~\& Doerschuk, P.},
{\rm 1998}:
{\rm The Humoral Immune-Response to Haemophilus-Influenzae 
Type-B -- A Mathematical-Model Based on T-Zone and Germinal 
Center B-Cell Dynamics},
{\it J. Theor. Biol.\/}
{\bf 194},
{\rm 341-381}.

\item
{\sc Siepmann, K., Skok, J., van Essen, D., Harnett, M.~\& Gray, D.},
{\rm 2001}:
{\rm Rewiring of CD40 is necessary for delivery of rescue 
signals to B cells in germinal centres and subsequent 
entry into the memory pool},
{\it Immunol.\/}
{\bf 102},
{\rm 263-272}.

\item
{\sc Skalak, R.~\& Chien, S.},
{\rm 1987}:
{\rm Handbook of Bioengineering. McGraw-Hill, New York et al.}

\item
{\sc Tew, J.~\& Mandel, T.},
{\rm 1979}:
{\rm Prolonged antigen half-life in the lymphoid follicles 
of specifically immunized mice},
{\it Immunology\/}
{\bf 37},
{\rm 69-76}.

\item
{\sc Tew, J.G., Wu, J., Qin,D., Helm, S., Burton, G.F.~\& Szakal, A.K.},
{\rm 1997}:
{\rm Follicular dendritic cells and presentation of antigen 
and costimulatory signals to B cells},
{\it Immunol. Rev.\/}
{\bf 156},
{\rm 39-52}.

\item
{\sc van Eijk, M.~\& de Groot, C.},
{\rm 1999}:
{\rm Germinal Center B-Cell Apoptosis Requires Both Caspase 
and Cathepsin Activity},
{\it J. Immunol.\/}
{\bf 163},
{\rm 2478-2482}.

\item
{\sc van Eijk, M., Medema, J.P.~\& de Groot, C.},
{\rm 2001}:
{\rm Cellular Fas-Associated Death Domain-Like IL-1-Converting 
Enzyme-Inhibitory Protein Protects Germinal Center 
B Cells from Apoptosis During Germinal Center Reactions},
{\it J. Immunol.\/}
{\bf 166},
{\rm 6473-6476}.

\item
{\sc Wedemayer, G.J., Patten, P.A., Wang, L.H., Schultz, 
P.G.~\& Stevens, R.C.},
{\rm 1997}:
{\rm Structural insigths into the evolution of an antibody 
combining site},
{\it Science\/}
{\bf 276},
{\rm 1665-1669}.

\item
{\sc Zhang, J., MacLennan, I.C.M., Liu, Y.J.~\& Land, P.J.L.},
{\rm 1988}:
{\rm Is rapid proliferation in B centroblasts linked to 
somatic mutation in memory B cell clones},
{\it Immunol. Lett.\/}
{\bf 18},
{\rm 297-299}.

\end{list}


\vspace*{5mm}
\subsection*{Acknowledgments}
I thank Tilo Beyer and Andreas Deutsch
for intense discussions and valuable comments.

\newpage


%
\begin{table}[ht]
\small
\begin{center}
\begin{tabular}{|l|c|c|c|} \hline
Parameter & symbol & physiological & assumed\\ \hline
Shape space dimension                    &$d_s$&&$4$\\
Width of Gaussian affinity weight function &$\Gamma$&${\it 2.8}$&\\
Lattice constant                         &$\Delta x$&&$10\micron$\\
Radius of GC                             &$R$&$220\micron$&\\
Number of seeder cells                   &$$&${\bf 3}$&\\
Diffusion constant for centroblasts      &$D_{CB}$&${\it 5}\frac{\micron^2}{hr}$&\\
Ratio of centroblast to centrocyte radius&$\frac{r_{CB}}{r_{CC}}$&${\bf 3}$&\\
Diffusion constant of signal molecules   &$D_{\rm sig}$&$200\frac{\micron^2}{hr}$&\\
Number of FDCs                           &$$&$20$&\\
Length of FDC arms                       &$$&$30\micron$&\\
Duration of phase of monoclonal expansion&$\Delta t_1$&${\bf 72\,hr}$&\\
Duration of optimization phase           &$\Delta t_2$&${\bf 48\,hr}$&\\
Rate of proliferation (2D)               &$p$&${\bf 1/(9hr)}$&\\
Maximal distance for CB proliferation    &$R_{\rm P}$&&$60\micron$\\
Mutation probability                     &$m$&${\bf 0.5}$&\\
Rate of differentiation signal production by FDCs&$s$&&$9/hr$\\
Rate of centroblast differentiation      &$g_1$&${\bf 1/(6hr)}$&\\
Rate of FDC-centrocyte dissociation      &$g_2$&$1/2hr$&\\
Rate of differentiation of selected centrocytes&$g_3$&$1/(7hr)$&\\
Probability of recycling of selected centrocytes&$q$&${\it 0.8}$&\\ 
Rate of centrocyte apoptosis             &$z$&$1/(7hr)$&\\
\hline
\end{tabular}
\caption[]{\sf Summary of all physiological and assumed 
magnitudes entering the 2D model. 
The parameters are classified in two basic categories:
Those with a direct {\it physiological} interpretation and
those which were {\it assumed} for conceptual reasons of the
model. The physiological parameters are shown in bold, if
they are known experimentally with a sufficient precision.
If the parameter has a physiological counterpart and is not
known experimentally but has been determined indirectly 
by the analysis of experimental data, the value is shown in italic.
Note, that all given rates are the physiological
ones which enter with an additional factor 
of $\ln(2)$ into the model. The symbols correspond
to those used in the text.}
\label{parameter}
\end{center}
\end{table}

\end{document}